\documentclass[%
 aip,
 jcp,
 amsmath,amssymb,
reprint,
]{revtex4-1}

\usepackage{graphicx}
\usepackage{dcolumn}
\usepackage{bm}
\usepackage[utf8]{inputenc}
\usepackage[T1]{fontenc}
\usepackage{mathptmx}

\usepackage{physics, amsmath, hyperref}
\usepackage{ulem} 
\usepackage{xcolor}
\usepackage{array, tabularx} 

\newcommand{\units}[1]{\,{\rm #1}}

\begin{document}

\title{The dark side of energy transport along excitonic wires:  \\
\large On-site energy barriers facilitate efficient, vibrationally-mediated transport through optically dark subspaces
}

\author{Scott Davidson}
\email{sd109@hw.ac.uk}
\affiliation{SUPA, Institute of Photonics and Quantum Sciences, Heriot-Watt University, EH14 4AS,  United  Kingdom}

\author{Amir Fruchtman}
\affiliation{Department of Materials, University of Oxford, Oxford OX1 3PH, United Kingdom}

\author{Felix A. Pollock}
\affiliation{School of Physics and Astronomy, Monash University, Clayton, Victoria 3800, Australia}

\author{Erik Gauger}
\affiliation{SUPA, Institute of Photonics and Quantum Sciences, Heriot-Watt University, EH14 4AS,  United  Kingdom}

\begin{abstract}
We present a novel, counter-intuitive method, based on dark state protection, for significantly improving exciton transport efficiency through `wires' comprising a chain of molecular sites with an intrinsic energy gradient. Specifically, by introducing `barriers' to the energy landscape at regular intervals along the transport path, we find that undesirable radiative recombination processes are suppressed due to a clear separation of sub-radiant and super-radiant eigenstates in the system. This, in turn, can lead to an improvement in transmitted power by many orders of magnitude, even for very long chains.  From there, we analyse the robustness of this phenomenon to changes in both system and environment properties to show that this effect can be beneficial over a range of different thermal and optical environment regimes. Finally, we show that the novel energy landscape presented here may provide a useful foundation for overcoming the short length scales over which exciton diffusion typically occurs in organic photo-voltaics and other nanoscale transport scenarios, thus leading to considerable potential improvements in the efficiency of such devices.
\end{abstract}

\maketitle

\section{Introduction}

\subsection{Background}
\label{sec:intro}

One of the fundamental challenges faced when designing, building or improving upon many modern technologies involves dealing with novel physical phenomena which are prevalent at nanoscopic length, time and energy scales. Unavoidable quantum mechanical effects can be particularly detrimental as we push towards ever smaller devices~\cite{Intro:transistor-tunnelling}. On the other hand, the same quantum mechanical effect also offer huge potential for technological advancements if we are able to understand and harness them correctly~\cite{Intro:quantum-computing-review, Intro:quantum-sensing-review, Intro:quantum-crypt-review}. One field in which the utilization of quantum effects could provide significant benefits is within devices for energy transport and storage. Natural light harvesting complexes, such as those found in photosynthetic bacteria, have received particular attention, due to the extremely high quantum yield observed in such systems~\cite{Intro:PhotosythesisMech} as well as the tantalizing prospect that the energies and length scales involved could potentially support non-trivial quantum mechanical phenomena. Regardless of whether or not these natural systems do genuinely make use of such phenomena, there exists a plethora of evidence which demonstrates that the same quantum mechanical effects could be hugely beneficial in artificially engineered light harvesting and energy transport devices~\cite{Intro:CoherentPhotoVoltaics, QT:EfficientPhoto, Intro:NanotechApplication}. 

\begin{figure}
    \includegraphics[scale=0.29]{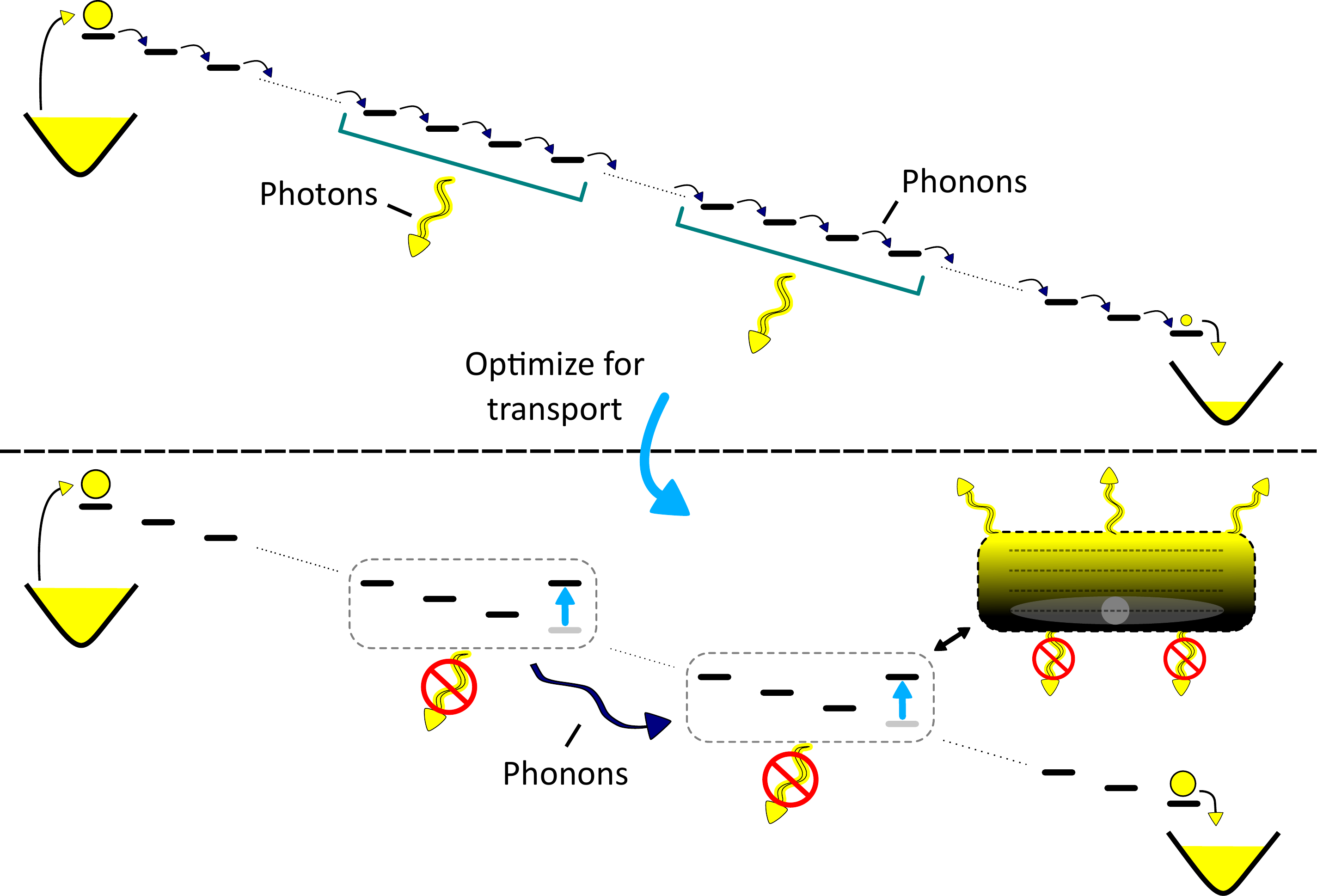}
    \caption{
    Significant enhancements in transport efficiency down an energy gradient can be achieved by adding regularly repeating `spikes' to the energy landscape, in order to prevent radiative loss. Essentially, an exciton then hops directly from the lower energy dark states of one `block' into the dark states of the adjacent block. The darkness of the lower states within each block arises from destructive interference in the collective light-matter coupling of its delocalised wavefunctions.
    }
    \label{fig:intro}
\end{figure}

The primary mechanism by which quantum effects may be beneficial in these devices is through the utilization of interference phenomena which, under the correct conditions, can induce either diffusive behaviour or localization effects. Furthermore, through the phenomenon of Dicke-superradiance~\cite{OQS:orig-superradiance-Dicke, OQS:superradiance-essay}, interference effects can even mitigate some of the detrimental effects of an external environment, which are unavoidable in any realistic device. These undesirable environmental effects are particularly prevalent in systems which harvest energy via the generation and subsequent transport of excitons (electron-hole pairs), where interactions with an ambient electromagnetic environment typically cause the excitons to recombine, radiating their energy back into the surroundings. Unlike the mechanisms causing non-radiative loss, radiative coupling is fundamentally unavoidable if we are to have resonant or incoherent F\"orster transport. In fact, in many organic photo-voltaics (OPVs), the small distances over which excitons typically travel before recombination are a decisive factor in the poor efficiency of these devices~\cite{QT:OPV-review}. Therefore, designing systems which can make use of such interference phenomena is highly desirable.

Previous works, both theoretical and experimental, have examined the use of super/sub-radiant interference effects as a possible means to enhance the performance of nanoscale energy transport and storage devices. In particular, dark-state protection has been proposed as a potential mechanism through which the mitigation of radiative losses may be achieved in both natural and artificial systems~\cite{QT:bio-inspired-dark-state-Creatore}. Most work in this space has focused on dimers\cite{ QT:QHE-ENAQT-Killoran,QT:DarkZhang, QT:Amir-photocell-prl,QT:super-cond-light-harv,QT:Rouse-strong-dark-state-dimers} and other small antenna structures;\cite{QT:DarkZhang,QT:chain-dark-state-photocell, OQS:Ratchets} however, this mechanism has recently been identified as playing an additional role in long-range energy transport within nanometer arrays of light harvesting antenna complexes~\cite{QT:long-range-LH2-array-experiment, QT:long-range-LH2-array-theory}.

The interplay between coherent interference effects and incoherent hopping processes also leads to the well-documented phenomenon of  Environment-Assisted Quantum Transport~\cite{QT:OriginalENAQT, QT:BuchleitnerReview} (ENAQT). Significant research efforts in this area have been devoted to understanding the circumstances under which environmental interactions are beneficial for transport~\cite{QT:MomentumRejuv, QT:OptimalNetworks, QT:OptimalNoisyQuantumWalks, QT:UniversalOriginENAQT} and also to demonstrating and utilizing these effects in artificial systems.\cite{QT:ion-chain-ENAQT, QT:EMG-molecule-junct-ENAQT, QT:waveguide-ENAQT}

In order to achieve directed energy transport, where net energy flow is from some initial `source' location to a spatially separated `drain', at which the transported energy is utilized, some form of imbalance is required within the transport medium. In charge transport scenarios, this is often achieved by applying a potential difference to allow charge carriers to flow from source to drain.~\cite{QT:Kouwenhoven-electron-transport, QT:Thijssen-charge-transport, Chen:GradientWire} Alternatively, in photosynthetic complexes (i.e.~natural exciton transport systems), it is believed that a net energy gradient is primarily responsible for efficient energy transport from the molecular antennae to the reaction center~\cite{QT:Kassal-photosynth-energy-funnelling}. In artificially engineered systems, similar funnelling mechanisms may also be possible through techniques such as strain-induced excitonic energy gradients~\cite{QT:exciton-energy-grad1, QT:exciton-energy-grad2, QT:exciton-energy-grad3} or chemical substitution in molecular chains~\cite{QT:molecular-energy-grad}.

In this Article, we theoretically study excitonic energy transport along a (molecular) chain which is subject to an external energy bias to allow net energy flow. Using this model, we then investigate whether there exists an on-site excitation energy configuration for this chain which provides improved transport performance over the (intuitively efficient) linear gradient. In doing so, we find a surprising class of energetic structures which, by utilizing the aforementioned collective effect of subradiance, can lead to systems in which transport is many orders of magnitude more efficient than the conventional linear chain (see Fig.~\ref{fig:intro}) if the dominant loss channel is radiative. We then proceed to analyse the physical mechanisms which allow this novel energy configuration to improve transport and find that the complex interplay between system eigenstate structure and vibrational and electromagnetic environments is of crucial importance. Despite this, the benefits of this energy configuration over the linear chain are visible over a broad range of both system and environment parameters which highlights the robustness of the effect. Finally, we explore the possibility of using the results described here as a building block template for significant efficiency improvements in arbitrarily long chains.

\subsection{The Model}
\label{sec:model}

\begin{figure}
    \includegraphics[scale=0.3]{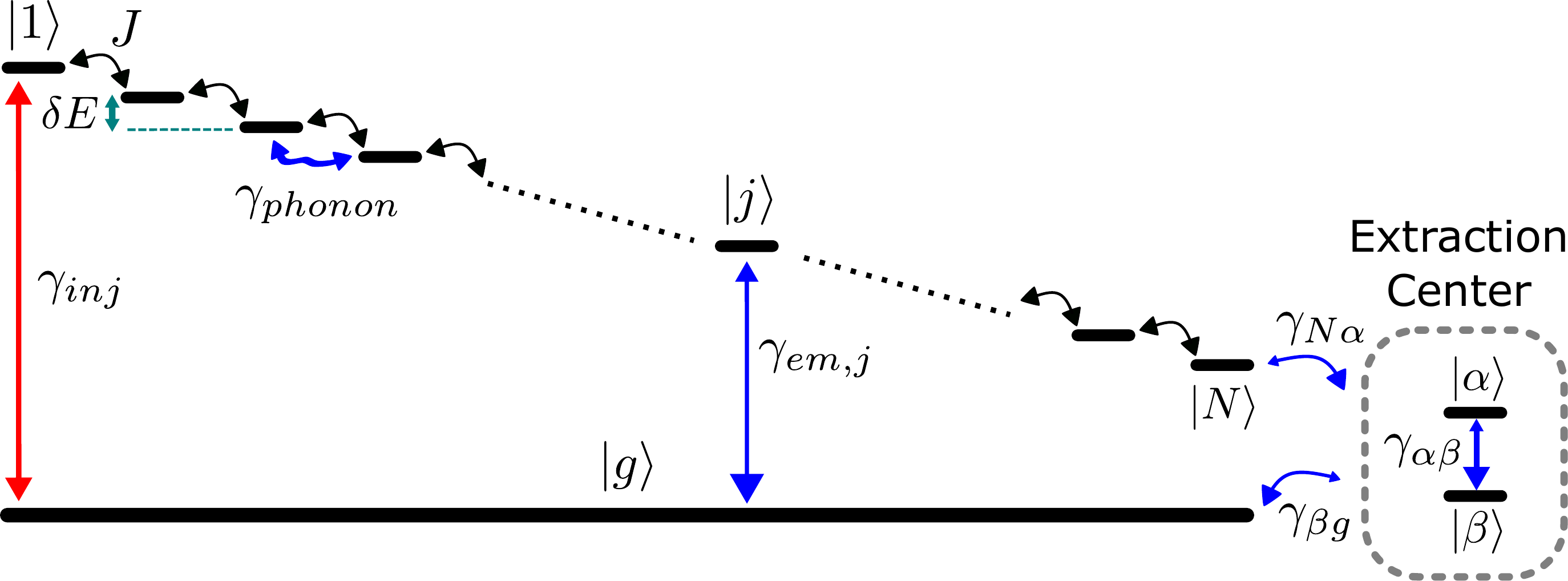}
    \caption{
    Depiction of the energetic structure and system-environment interactions included in the open quantum systems model described in the main text. Red and blue arrows represent processes with thermal baths at temperatures $T_{hot} = 6000\units{K}$ and $T_{cold} = 300\units{K}$ respectively. (See Table~\ref{table:param-values} for other parameter values used)
    }
    \label{fig:system}
\end{figure}

In this section, we construct an open quantum systems model to describe excitonic energy transport along a biased chain (see Fig.~\ref{fig:system}). The `system' is comprised of $N$ sites where the on-site excitation energies form a linear energy gradient, with energy difference $\delta E$ between neighbouring sites, and each site connected to its nearest neighbours via a (resonant F\"orster) hopping interaction with strength $J$. We also include an overall ground state, in which there are no excitations in the chain, and assume that there is never more than a single excitation in the chain at a time. The Hamiltonian for the chain system is given by
\begin{eqnarray}
    \label{eq:H-chain}
    \hat{H}_\text{chain} &=& \sum_{j=1}^{N} \left( \varepsilon_e - j \cdot \delta E \right) \outerproduct{j}{j}  \nonumber \\
    &+& \sum_{|i-j|=1} \frac{J}{2} \outerproduct{j}{i} + \varepsilon_g \outerproduct{g}{g} ~,
\end{eqnarray}
where $\ket{j}$ describes a state with an excitation on the $j$-th site, $J$ is the nearest neighbour coupling strength, and $\varepsilon_g$ and $\varepsilon_e$ set the energy difference between the ground state $\ket{g}$ and the excited state manifold. Throughout most of the main text we will focus on chains of length $N=20$; however, as shown in Sec.~\ref{sec:group-opt}, our results also hold for much longer chains. All other parameter values are provided in Table~\ref{table:param-values}.

\begin{table}
    \begin{tabularx}{0.45\textwidth}{
        | >{\centering\arraybackslash}X 
        | >{\centering\arraybackslash}X |
        | >{\centering\arraybackslash}X 
        | >{\centering\arraybackslash}X |
        }

        \hline
        \textbf{Parameter} & \textbf{Value} & \textbf{Parameter} & \textbf{Value} \\
        \hline
        $N$ & 20 sites                                      &   $\gamma_{em, j}$ & 1e-3 eV                      \\
        $\delta E$ & 0.05 eV                                &   $\gamma_{inj}$ & 2.3e-3 eV                      \\
        $J$ & 0.1 eV                                        &   $\gamma_{phonon}$ & 1.4e-3 eV                   \\
        $\varepsilon_g$ & 0 eV                              &   $\Gamma$ & 2e-2 eV                              \\
        $\varepsilon_e$ & 1.65 eV                           &   $\omega_0$ & $\sqrt (\delta E^2 - \Gamma^2)$    \\
        $\varepsilon_\alpha$ & 0.2 eV                       &   $\gamma_{N\alpha}$ & 1.05e-3 eV                 \\
        $\varepsilon_\beta$ & 0.5 eV                        &   $\gamma_{\beta g}$ & 1.3e-3 eV                  \\
        $T_\text{hot}$ & 6000 K                             &   $\gamma_{nr}$ & 0 eV                            \\
        $T_\text{cold}$ & 300 K                             &   $T_\text{ph}$ & 300 K                           \\
        \hline
    \end{tabularx}
    \caption{Default parameter values used in the model described in Sec.~\ref{sec:model} for all calculations unless explicitly stated otherwise. Some of these parameters choices are inspired by prior related work.\cite{QT:Amir-photocell-prl,OQS:Amir-thesis,QT:bio-inspired-dark-state-Creatore} However, throughout this study we aim to focus on general physical properties and mechanisms, rather than a particular material system.
    }
    \label{table:param-values}
\end{table}

We also account for ambient vibrational and electromagnetic environments, which would typically surround such an energy transport system, in any practical scenario. Assuming weak system-environment coupling for simplicity and computational tractability, we use a Pauli Master Equation~\cite{OQS:BreuerPetruccione} (PME) to describe energy transport through our chain system. The PME takes the general form
\begin{equation}
    \label{eq:pme}
    \partial_t P_n = \sum_m \left[ W_{nm} P_m(t) - W_{mn} P_n(t) \right] ~,
\end{equation}
where $P_n$ is the population of the $n$-th eigenstate. The matrix $\mathbf{W}$, which determines the transitions rates between eigenstates, takes the form
\begin{equation}
    \label{eq:W-mat}
    W_{nm} = \sum_\mu S_\mu (\omega_{mn}) \bra{\phi_m} A_\mu \ket{\phi_n} \bra{\phi_n} A_\mu \ket{\phi_m} ~,
\end{equation}
where $\omega_{mn} \equiv \varepsilon_m - \varepsilon_n $ denotes the energy difference between eigenstates $\ket{\phi_n}$ and $\ket{\phi_m}$, $\mu$ denotes each of the different system-environment interactions and $A_\mu$ and $S_\mu$ are the relevant system operator and density of states for each interaction respectively. The key results presented here were also verified using an equivalent Bloch-Redfield master equation, however, due to numerical efficiency considerations, the PME was used for the majority of the calculations.

The vibrational environment is modelled as a local phonon bath coupled to each system site with an interaction operator of the form
\begin{equation}
    \label{eq:phonon-op}
    \hat{A}_{j,\text{ vib}} = \outerproduct{j}{j} ~.
\end{equation}
The primary effect of this thermal environment is to induce phonon-mediated transitions between the system eigenstates at rates determined by the phonon spectral density $S(\omega)$, for which we use an ohmic spectrum with Drude-Lorentz cutoff
\begin{equation}
    \label{eq:Sw-DL}
    S_{DL}(\omega) = |\omega| \frac{\pi \Gamma \gamma_{phonon}}{\Gamma^2 + (|\omega| - \omega_0)^2} \left[n_{BE}(|\omega|, T_{ph}) + \Theta(\omega)\right] ~,
\end{equation}
where $\gamma_{phonon}$ controls the overall system-phonon coupling strength, $\Gamma$ is the spectral width and $\omega_0$ is a high-frequency cutoff. The Bose-Einstein distribution $n_{BE}$, at temperature $T_{ph}$, describes the thermal occupancy of each phonon mode and the asymmetry introduced by the Heaviside step-function $\Theta$ accounts for the dominance of phonon emission over absorption events in a realistic finite temperature vibrational bath. This asymmetry leads to downward energy transitions (i.e.~phonon emission by the system) being favoured so that, in the absence of other environmental effects, the steady state of the system is dominated by the lowest energy eigenstates (i.e.~excitons are funneled towards the bottom of the chain). In all of the discussion that follows, and unless otherwise stated, the phonon spectrum parameters in Eq.~\eqref{eq:Sw-DL} are chosen to favour transport along a linearly biased chain. This means that we choose $\Gamma$ and $\omega_0$ such that the peak of the phonon spectrum (without the term in square brackets) is at $\omega_\text{max} = \delta E$. (Specifically, we fix $\Gamma$ and then let $\omega_0^2 = \delta E^2 - \Gamma^2$.) This means that phonon processes are most efficient at funnelling excitations to the bottom of the chain when the system eigenenergies are equally spaced, as is the case for the linear gradient (up to some finite-size effects). This allows the chain with linear on-site energy gradient, which we use as a benchmark throughout this work, to make best use of the phonon environment for efficient transport and therefore provides a fair comparison with the alternative on-site energy configurations discussed in the rest of the paper.

For the electromagnetic environment we assume that, due to the nanoscopic scale of the system, each of the sites in the system couple to the ambient electromagnetic field with the same phase. This `collective' optical coupling leads to the well-documented emergence of super/sub-radiant system eigenstates~\cite{Intro:QO-for-beginners-Ficek2014, OQS:infinite-chain-super-rad, OQS:waveguide-collective-effects, OQS:surface-plasmon-collective-effects}, where the super-radiant states exhibit an enhanced coupling to the field, leading to faster optical processes, while the sub-radiant states couple to the field extremely weakly and therefore possess extremely long optical lifetimes. As will be relevant in Sec.~\ref{sec:group-opt}, we note here that the validity of the collective optical coupling description relies only on the wavelength of the electromagnetic radiation being much greater than the typical localization length of the individual eigenstates, rather than the typical length scale of the entire system. Since the former is significantly shorter in this system (due to the exponential localization of eigenstates resulting from the energy detuning between sites) the collective coupling regime is valid regardless of the length of the chain. The influence of this ambient electromagnetic field on the system is described by
\begin{equation}
    \label{eq:optical-op}
    \hat{A}_{em} = \sum_{j=1}^N \left( \outerproduct{j}{g} + \outerproduct{g}{j} \right) ~.
\end{equation}
The spectral density for this electromagnetic interaction is taken to be flat around the relevant frequencies, reflecting the typically Markovian nature of the photon bath, and takes the form
\begin{equation}
    \label{eq:Sw-flat}
    S_{\text{flat}}(\omega) = \gamma_{em} \left[ n_{BE}(\omega, T_{\text{cold}}) + \Theta(\omega) \right] ~,
\end{equation}
where $\gamma_{em}$ dictates the overall coupling to the field (which is the same for all sites) and $n_{BE}$ and $\Theta$ have the same meanings as in Eq.~\eqref{eq:Sw-DL}.

Additionally, we also include several other phenomenologically motivated, and environmentally induced, processes in order to model other aspects of a realistic energy transport system. Firstly, following Ref.~[\onlinecite{QT:QHE-Dorfman-Scully}], we include a `extraction center' (EC) which mimics the utilization of energy quanta after they have been successfully transported along the chain, analogous to the reaction centre in photosynthesis or a load in an electric circuit. For this purpose, we include two extra sites in the system's Hilbert space, indexed by $\alpha$ and $\beta$ (Fig.~\ref{fig:system}), which represent the excited and ground states of the EC. Our total system Hamiltonian is therefore given by
\begin{equation}
    \hat{H}_s = \hat{H}_\text{chain} + \varepsilon_\alpha \outerproduct{\alpha}{\alpha} + \varepsilon_\beta \outerproduct{\beta}{\beta} \, ,
\end{equation}
where $\varepsilon_\alpha$ and $\varepsilon_\beta$ are the energies of the EC excited and ground states respectively. We also include three extra system-environment interactions which allow population to flow through the EC. The operators describing these three processes are of the form
\begin{equation}
    \label{eq:generic-env-op}
    \hat{A}_{ab} = \outerproduct{a}{b} + \outerproduct{b}{a} ~,
\end{equation}
where the pairs of system indices (represented by $a$ and $b$) are ($N, \alpha$), ($\alpha, \beta$) and ($\beta, g$), respectively. The frequency spectra for each of these processes is taken to be that of Eq.~\eqref{eq:Sw-flat} but with the rate $\gamma_{em}$ replaced by the relevant rate $\gamma_{ab}$ for each of the three respective processes. Finally, we include an injection process which describes the initial generation of an exciton at the top of the chain (e.g. by F\"orster-like excitation via a proximal dipole). The operator for this process takes the form
\begin{equation}
    \label{eq:inj-op}
    \hat{A}_{inj} = \outerproduct{g}{1} + \outerproduct{1}{g} ~,
\end{equation}
and the spectrum is that of Eq.~\eqref{eq:Sw-flat} with rate $\gamma_{inj}$ and temperature $T_{cold}$ replaced by $T_{hot} = 6000$\units{K}. This hotter bath temperature increases the rate of absorption processes which populate site $\ket{1}$ compared with those which work in the opposite direction.\footnote{Note - we could also have realised a one-way injection process to the first site instead of coupling to a heat bath. However, with appropriate re-scaling of parameters to match the excitation rate in the two cases, this does not affect our conclusions.} Altogether, the sum over $\mu$ in the rate matrix $\mathbf{W}$ [Eq.~\eqref{eq:W-mat}] of our PME model encompasses each of the system interaction operators ($\hat{A}$'s) described above.

Before investigating whether or not a simple linear gradient in the on-site energies of our system is optimal for transport, we must first define a useful measure of transport efficiency. To this end, we focus on steady state transport in which excitations are continuously generated at the top of the chain and, following references~\cite{QT:QHE-Dorfman-Scully, QT:QHE-ENAQT-Killoran, QT:Amir-photocell-prl}, we use the power output from the extraction center, defined via
\begin{align}
    I &= e \gamma_{\alpha\beta} P_\alpha ~, \notag \\
    V &= \varepsilon_\alpha - \varepsilon_\beta + k_b T_\text{cold} \log ( P_\alpha / P_\beta ) ~, \notag \\
    \label{eq:power}
    \mathcal{P} &= I V 
\end{align}
as the primary measure of transport efficiency in our model system. We treat the EC as a effective `black box' and therefore define the power output from the chain as the maximal power achievable by varying the rate $\gamma_{\alpha\beta}$. Despite focusing on power output in the main text, we note that the results presented in the remainder of this paper are generally applicable to other measures of transport efficiency; as shown by an analysis of the steady state exciton current through our system in Appendix~\ref{apdx:ss-current}.

\section{Optimizing Transport}

\subsection{Sensitivity Analysis}
\label{sec:param-sens}

\begin{figure}
    \includegraphics[scale=0.29]{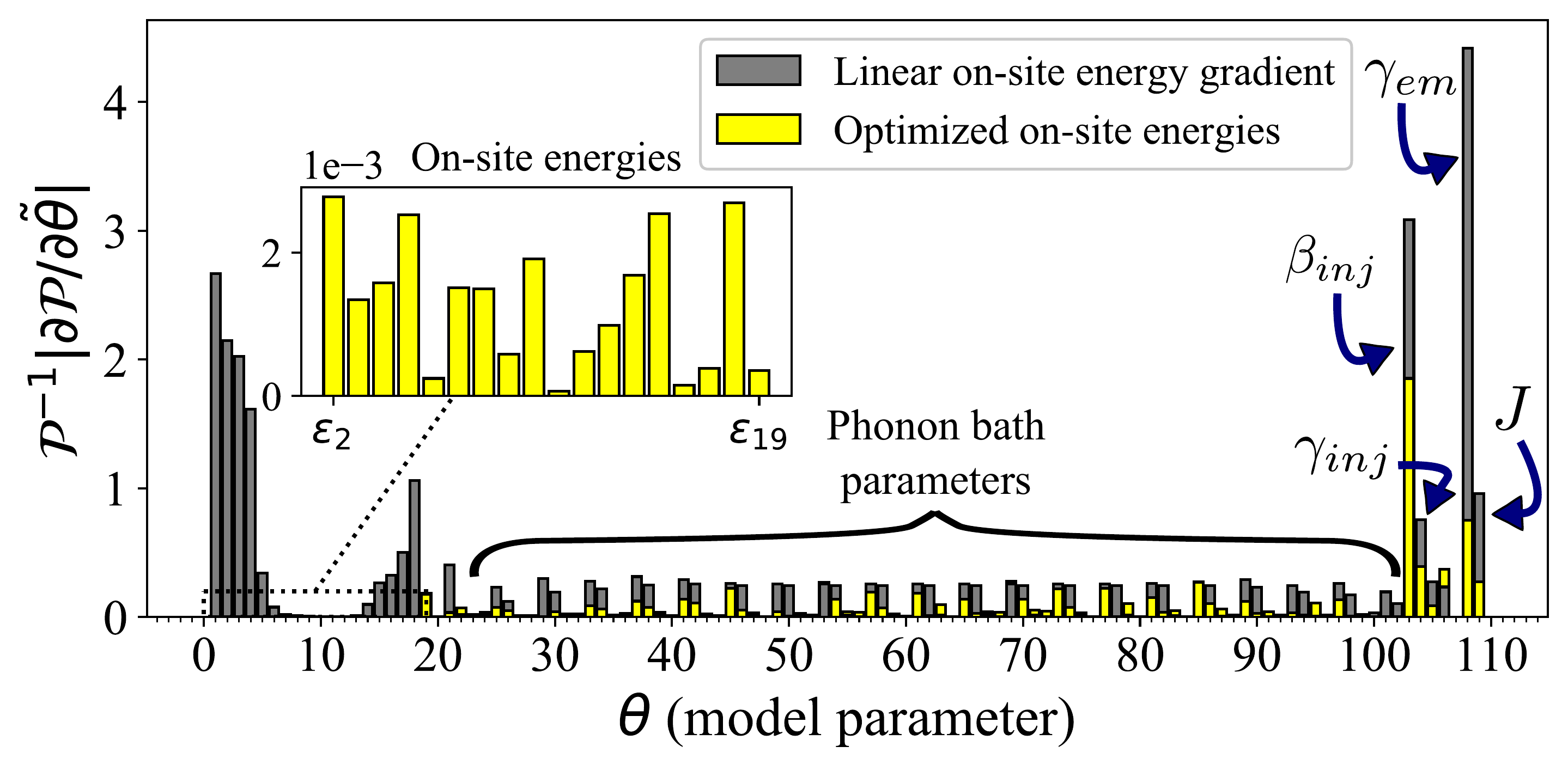}
    \caption{
	Parameter sensitivity analysis w.r.t. output power ($\mathcal{P}$) for the model shown in Fig.~\ref{fig:system} before (grey) and after (yellow) on-site energy optimization. Coupling to the ambient electromagnetic field is the biggest factor in achieving maximal power output for the linear energy gradient. After on-site energy optimization, we see a significant reduction in sensitivity to both on-site energy changes and $\gamma_{em}$ indicating a partial decoupling from the optical environment via the formation of dark states (see Sec.~\ref{sec:mechanism}).
    }
    \label{fig:param-sens}
\end{figure}

As a means of investigating which parameters should be tuned most carefully in order to improve the transport performance of the model described in the previous section, we begin by carrying out a parameter sensitivity analysis with respect to the output power [Eq.~\eqref{eq:power}]. Adapting the approach used in Ref.~[\onlinecite{SD-paper1}], we analyse our model using dimensionless (\textit{log}) parameters ($\tilde{\theta} \equiv \log \theta$; where $\theta$ denotes the bare model parameter) to allow for fair comparison between quantities with different physical dimensions. Upon doing so, as shown in Fig.~\ref{fig:param-sens}, we find that in the linear chain configuration, the most important consideration for achieving efficient exciton transport is the coupling to the ambient electromagnetic field ($\gamma_{em}$). Transport is also relatively sensitive to the injection process temperature and coupling strength (parameterized by $\beta_{inj}$ and $\gamma_{inj}$ respectively) and the on-site energies near the top of the chain. The importance of $\gamma_{em}, \beta_{inj}$ and $\gamma_{inj}$ makes intuitive sense, since `obtaining excitons to be transported' and `not losing those excitons during transport' are both trivial requirements for power generation at the EC. The sensitivity to only the first few on-site energies becomes clear when considering that, for the linear chain, most of the population as been lost to radiative decay near the top of the chain (see Fig.~\ref{fig:eigenbasis} in Sec~\ref{sec:mechanism}). Therefore, the energies in the middle of the chain are largely irrelevant. However, for the small fraction of population which does reach the end of the chain, the remaining on-site energies at the bottom of the chain become important for maximizing power output. In comparison, the sensitivity of transport performance to changes in other model parameters is negligible (see Table~\ref{tab:param-sens-map} in Appendix~\ref{apdx:param-table} for a full list of x-axis parameters in Fig.~\ref{fig:param-sens}). 

In practice, completely decoupling the system from the electromagnetic field (i.e. reducing $\gamma_{em}$ to avoid exciton recombination) is impossible, as it would impede the possibility of capturing excitons at the top of the chain and also result in vanishing hopping interaction strength $J$, preventing transport altogether.\footnote{Note that our simplified model does not explicitly include the link between the hopping strength $J$ and the radiative loss rate $\gamma_{em}$. The connection is straightforward in atomic systems (see e.g.~Ref.~[\onlinecite{OQS:Will-GuideSlide}]), but more complex in typical molecular settings.\cite{QT:Rouse-strong-dark-state-dimers}} For these reasons, throughout this paper (other than Sec.~\ref{sec:var-envs}) we assume that this coupling to the electromagnetic field is fixed. Going further, we also assume that it would be practically difficult to independently alter or control the various other parameters in our model pertaining to any of the system-environment interactions. Instead, we focus on investigating whether changes in the system parameters, specifically the on-site excitation energy of each site, can have a meaningful effect on exciton transport performance.

The results of this sensitivity analysis also demonstrate that, by optimizing only the on-site energies in the system, it is possible to achieve reduced sensitivity to many of the other parameters characterising the external environment, thus making transport more robust to perturbations in general (as shown by the overall reduction in most of the derivatives in Fig.~\ref{fig:param-sens} after on-site energy optimization).

\subsection{Optimized Energy Configurations}
\label{sec:optim}

We begin our investigation of optimal energy landscapes by performing a numerical optimization of the on-site energies in our system with respect to the power output defined in Eq.~\eqref{eq:power}. In order to maintain an intrinsic energy gradient from injection site to EC in our model, we fix the energies of the first and last sites of the chain (with a total energy difference $\varepsilon_1 - \varepsilon_N = N \cdot \delta E$) and focus on finding the best possible configuration for the $N-2$ other energies. Therefore, accounting for the optimization of rate $\gamma_{\alpha\beta}$, this is an $N-1$ parameter optimization problem. As shown explicitly in Appendix~\ref{apdx:local-opt}, as $N$ is increased, the complexity of the parameter space landscape increases commensurately. This leads to the presence of a large number local optima with similar efficiencies and qualitative features but quantitatively different energy landscapes. With this in mind, it becomes far more productive to look for commonalities between these different local optima rather than searching for a single definitive global optimum. To this end, we approach this optimization problem using a combination of local optimization algorithms including simplex-based~\cite{Nelder-Mead-alg} and gradient-based~\cite{numerical-optim-BFGS} methods. We further increase our ability to sample a range of local optima by starting at a number of randomly disordered configurations in which the on-site energies are slightly perturbed from the perfect linear gradient. (See Appendix~\ref{apdx:local-opt} for details.)

\begin{figure}
    \includegraphics[scale=0.43]{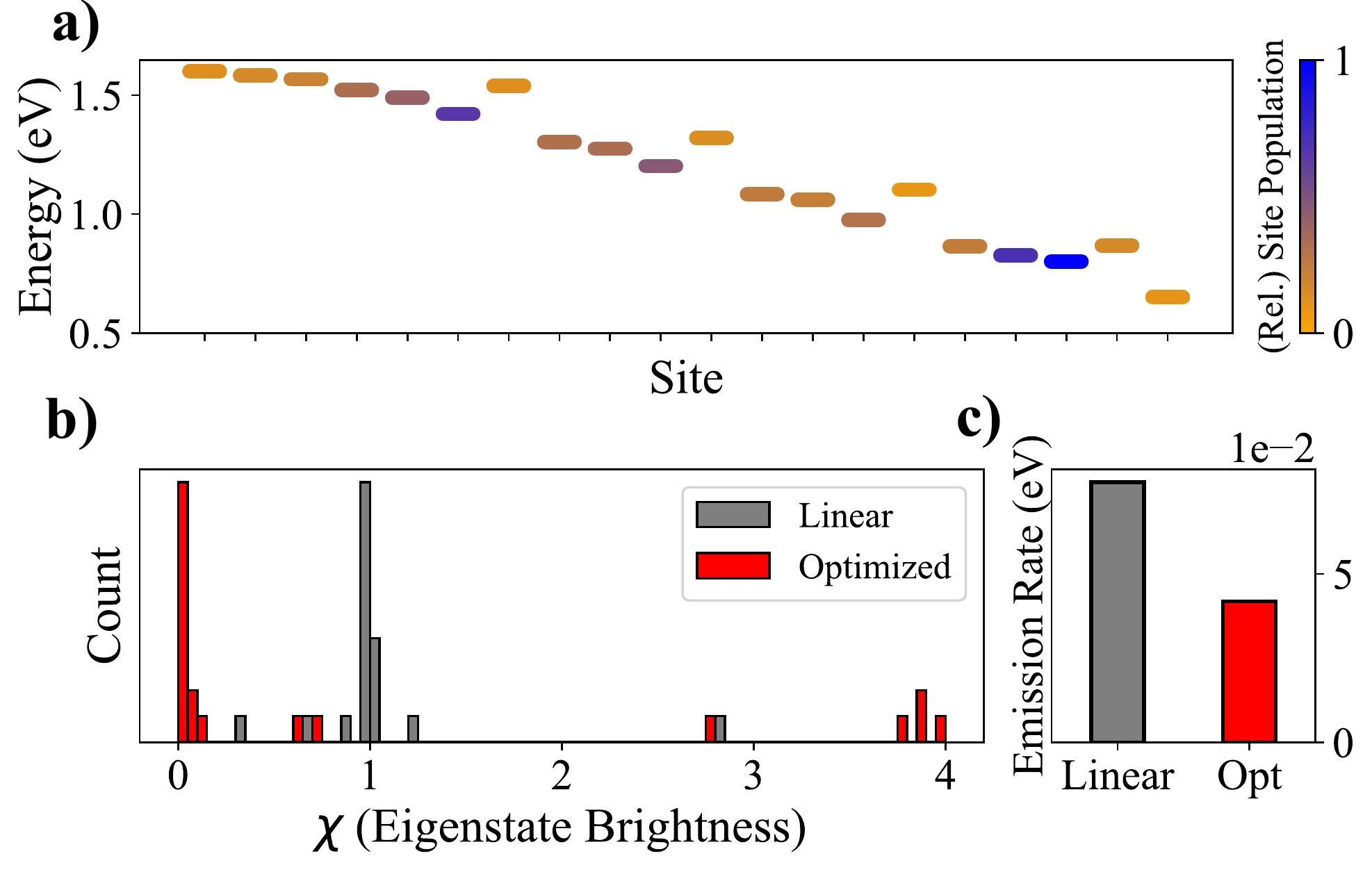}
    \caption{
    \textit{\textbf{a)}} Typical example of a local optimum with regularly spaced energy `spikes'. \textit{\textbf{b)}} Evidence of bright/dark state separation after on-site energy optimization and \textit{\textbf{c)}} the resulting reduction in steady state emission from the chain, given by $\sum_n P_n \chi_n$ where $P_n$ is the population of the $n$-th eigenstate of the system and $\chi$ is defined in Eq.~\eqref{eq:brightness}. Here, we set $\varepsilon_1 = 0 \units{eV}$ and $\varepsilon_N = -0.95 \units{eV}$; for all other parameter values see Table~\ref{table:param-values}.}
    \label{fig:example-opt}
\end{figure}

Upon performing this numerical optimization procedure and inspecting the energy configurations of the best performing chains, we build on the observations in Ref.~[\onlinecite{OQS:Amir-thesis}] and find that they each possess distinctive `spikes' in the on-site energies at regular points along the chain (see Fig.~\ref{fig:example-opt}a) and that these energy configurations are significantly more efficient than the linear gradient ($\sim$ 27 times more efficient for the 20 site chain shown here but can be many orders of magnitude higher in certain parameter regimes - see Sec.~\ref{sec:rad-vs-nonrad}). Furthermore, by looking at the brightness ($\chi$) of each system eigenstate $\ket{\phi_n}$, defined as
\begin{equation}
    \label{eq:brightness}
    \chi_n = |\bra{g} \hat{A}_{em} \ket{\phi_n}|^2 ~,
\end{equation}
we notice that these optimized chains exhibit a dramatic separation into super-radiant (bright) and sub-radiant (dark) states compared with the linear gradient configuration (see Fig.~\ref{fig:example-opt}b). This results in a reduced exciton recombination rate which contributes to the large efficiency enhancements.

These regular spikes in energy along the optimized chain intuitively act as potential barriers for the energy quanta in the chain and should, if anything, \textit{inhibit} transport by forcing the excitons to bypass these barriers if transport is to be successful. In contrast, we find that the inclusion of these barriers allows transport to proceed through a sequence of optically dark states while making use of phonon interactions to avoid populating the bright states.The utilization of sub-radiant states for improved transport has previously been explored in both photosynthetic~\cite{QT:long-range-LH2-array-theory} and quantum optics~\cite{QT:Plankensteiner-subradiant-rings} settings. In Ref.~[\onlinecite{QT:long-range-LH2-array-theory}], the on-site energies were subject to random static disorder, and variations in inter-site coupling strength were instead utilized in order to leverage dark-state protection for efficient transport. Alternatively, in Ref~[\onlinecite{QT:Plankensteiner-subradiant-rings}], the dipole orientations of the different sites in a ring system were varied, and the spatial distribution of the electric field from each eigenstate was investigated for dark and bright states separately. It was found that sub-radiant states in one ring system couple strongly to similar sub-radiant states in an adjacent ring system, which suggests the possibility of longer range transport through a sequence of dark states on each ring. There also exist a plethora of studies concerned with sub-radiance in the field of quantum optics, e.g.~focusing on state preparation and control schemes for manipulating radiative decay processes~\cite{QT:Plankensteiner-subradiant-rings, QO:Scully-subradiance, QO:Zoller-distant-subradiant-bell-states, QO:Selective-radiance-photon-storage}. By and large these show little consideration for vibrational and non-radiative effects. By contrast, the mechanisms underlying our findings make explicit use of inter-site energy detuning and the presence of both (finite temperature) vibrational and electromagnetic environments in order to achieve dark-state enhanced transport. In the next section we explore the complex system-environment interplay which leads to this novel `environment-assisted' phenomenon.

\subsection{Mechanisms Underlying Enhancement}
\label{sec:mechanism}

\begin{figure*}
    \includegraphics[scale=0.47]{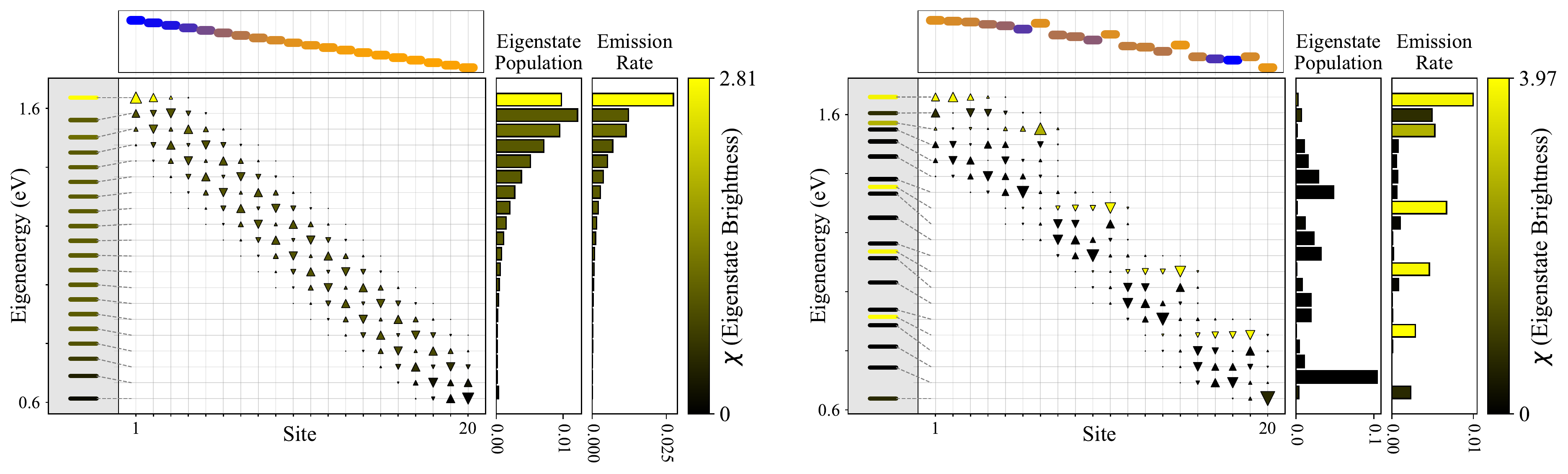}
    \caption{
    Comparison of the eigenbasis properties of chains with a linear energy gradient (\textit{left}) and those (\textit{right}) featuring on-site energy spikes. 
    \textit{\textbf{Left}} - Linear energy gradient leads to a single bright (dark) state at the top (bottom) of the chain with the majority of the eigenstates having roughly unit brightness (i.e.~no strongly sub-radiant or super-radiant behaviour) and progressively smaller populations upon moving towards the lower end of the chain (due to continuous radiative losses). The top panel shows the site-basis energies (with the same colourscale as Fig.~\ref{fig:example-opt}) and the left-most panel shows the eigenenergies of the Hamiltonian with a dotted grey line connecting each eigen-energy to the corresponding eigenstate. The large central panel shows the site-basis components of each eigenstate, with the size and (up/down) orientation of the triangles representing the magnitude $|\bra{\phi_n}\ket{j}|^2$ and relative phase ($=\pm 1$), respectively, of each component within the eigenstate. The horizontal bar plots show the population $P_n$ and steady state emission rate $\chi_n$ [Eq.~\eqref{eq:brightness}] from each of the eigenstates in the central panel. 
    \textit{\textbf{Right}} - The same plot format for a chain with optimized energy configuration featuring on-site energy spikes (see Fig.~\ref{fig:example-opt}). Here we see distinctive blocks of localized eigenstates with a single, minimally-populated bright state at the top of each block, and several well-populated dark states at lower energies within each block. 
    }
    \label{fig:eigenbasis}
\end{figure*}

In order to analyse the mechanisms by which these energetic barriers enhance transport, it is useful to look at the structure of the eigenenergies and eigenstates of the optimized chain. Figure~\ref{fig:eigenbasis} compares these eigenbasis properties for a linear energy gradient and the optimized energy configuration containing energy `spikes'. In the latter, the eigenstates clearly separate into discrete, localized `blocks' in the site-basis (up to some boundary effects) with each block containing an optically bright state as its highest energy state and several dark states at lower energies. Although the specific optimum analysed here contains blocks which are localized over precisely four sites, the general mechanism examined in this section holds for other similar block sizes (as shown in various appendices). The precise number of sites in each block is highly dependent on which local maximum the numerical optimization procedure converges to (see Appendix.~\ref{apdx:local-opt})

\begin{figure}
    \includegraphics[scale=0.48]{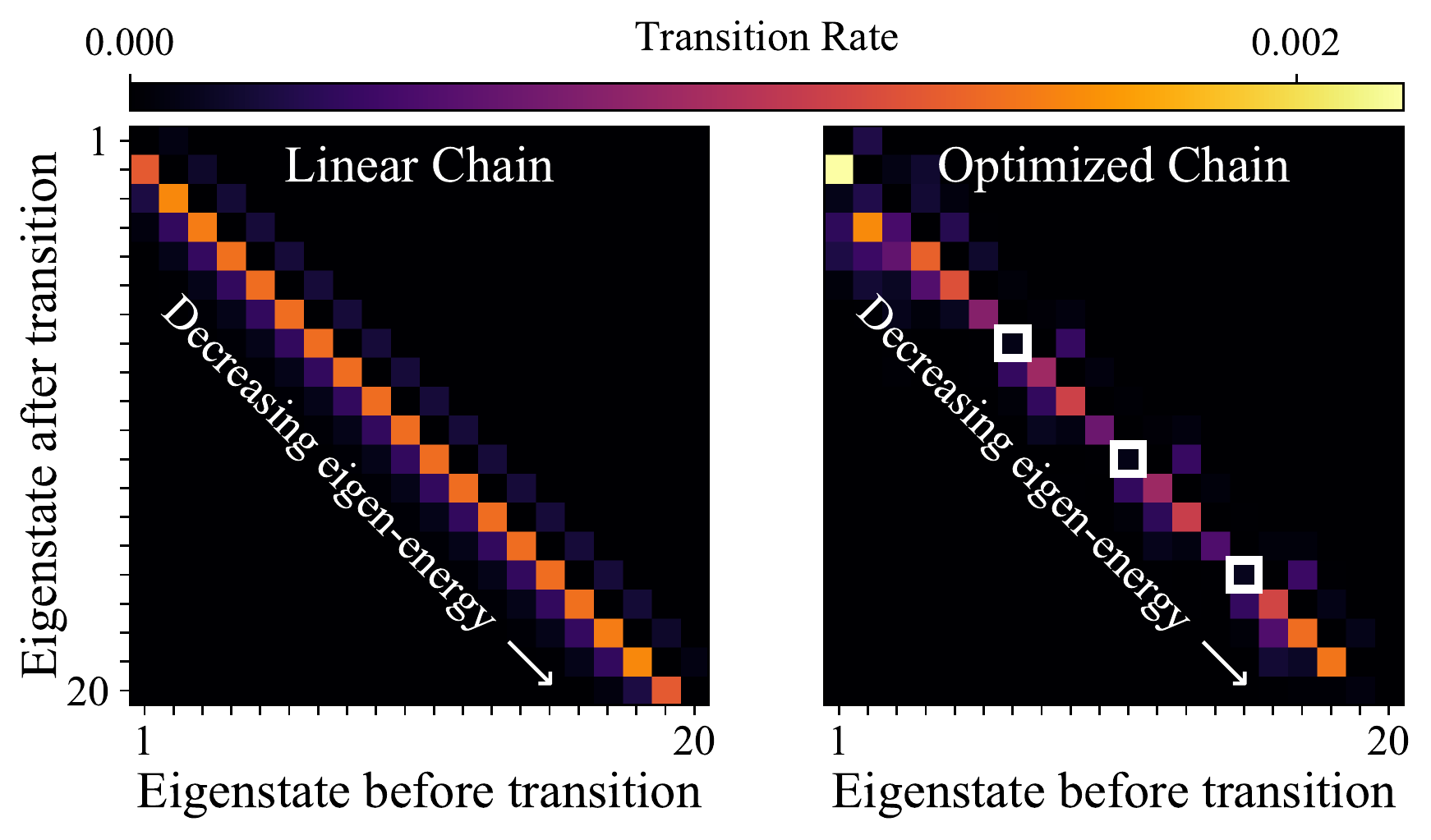}
    \caption{
    Phonon transition rates between system eigenstates before ({\it left}) and after ({\it right}) on-site energy optimization. The lack of transitions which populate bright states (highlighted by solid white squares) contribute to the significant efficiency enhancements seen in the optimized chains.
    }
    \label{fig:phonon-rates}
\end{figure}

The eigenstate population and emission rate panels demonstrate that, in the optimized case, the excitons spend the majority of their time during transport in the states at the bottom of each eigenstate block and that these states have negligible steady state optical emission rates, thereby helping to avoid exciton recombination events. The concentration of population in these lower energy dark states is achieved partly by the dominance of phonon emission events at finite temperature but is also crucially boosted by the noticeable `left-right' separation of site basis support between the bright and dark states in each eigenstate block. In our Pauli master equation formalism, the rates of the phonon-mediated transitions between eigenstates due to the local phonon bath at site $\ket{i}$ are given by $W_{nm, i} = S_{DL}(\omega_{mn}) |c_{mi}|^2 |c_{ni}|^2$ where $S_{DL}$ is the phonon spectral density (Eq.~\ref{eq:Sw-DL}) and $c_{mi} \equiv \braket{\phi_m}{i}$ is the overlap between eigenstate $\phi_m$ and site $i$ (see Appendix~\ref{apdx:phonon-rates} for proof). This tells us that, for a flat density of states, eigenstate pairs which share regions of spatial localization (i.e.~are `closer' to each other) will have stronger transition rates. This picture is complicated by the structured phonon bath spectrum used in our model (Eq.~\eqref{eq:Sw-DL}); however, for eigenstates which are sufficiently close in energy, this analysis still holds. This results in phonon-mediated transitions between the lowest energy dark state in one eigenstate block, and the highest energy \textit{dark} state in the next block being more likely to occur than transitions which populate the intermediate (in energy) \textit{bright} state. This effect is clearly demonstrated in Fig.~\ref{fig:phonon-rates}, which shows the phonon-mediated transition rates between all pairwise combinations of eigenstates and highlights the lack of transitions which populate bright states.

In summary, by tuning the site-basis energies in the system, we are able to modify the system-environment couplings between both the vibrational and electromagnetic environments in such a way as to generate a small number of minimally populated, optically bright states which allows transport to proceed far more efficiently through the remaining (optically dark) states.

Since the mechanism presented here is reliant on the combination of a number of independent physical effects, it would be tempting to assume that these novel energy configurations are only beneficial when each of the independent parameters are finely tuned to complement each other. However, in the next two sections, we will show that these regularly `spiked' energy landscapes are in fact more efficient than a chain with linear energy gradient over a significant range of parameter values.

\section{Varying System Parameters}
\label{sec:dE-and-J}

\begin{figure}
    \includegraphics[scale=0.43]{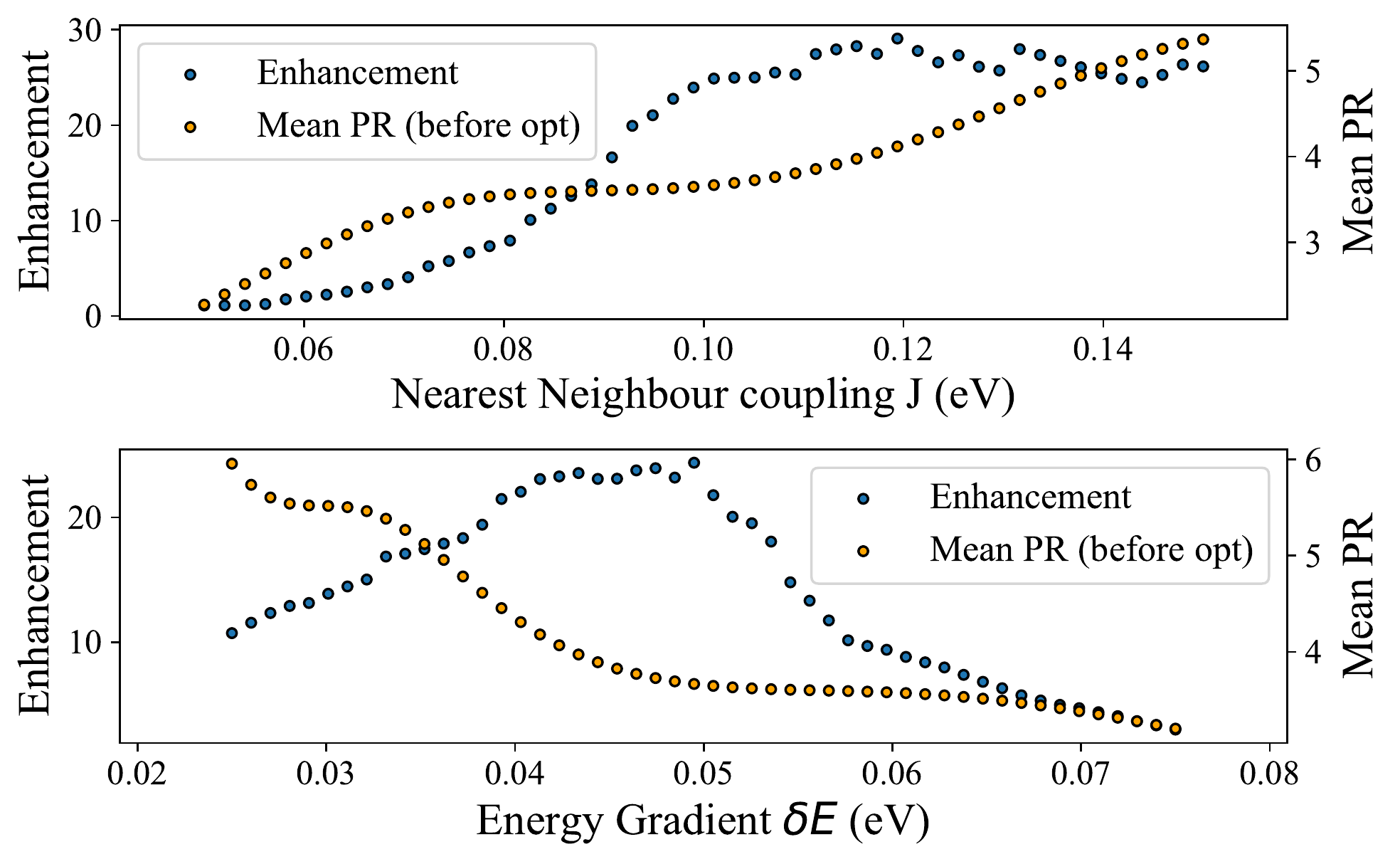}
    \caption{
    \textit{Top} - Maximum enhancement factor found by performing the optimization procedure described in Sec.~\ref{sec:optim} at each value of $J$ and the average eigenstate Participation Ratio (PR) before optimization (given by $\langle \langle \sum_i \left[ \braket{\phi_n}{i}\!\braket{i}{\phi_n} \right]^{-2} \rangle \rangle $; where $\ket{i}$ is a site-basis state, $\ket{\phi_n}$ is a system eigenstate and $\langle \langle \hdots \rangle \rangle$ denotes the average over eigenstates). \textit{Bottom} - Maximum enhancement factor found, along with average PR, when performing the aforementioned optimization procedure for a range of initial energy gradients instead of $J$ values. In both cases enhancements are largest when eigenstates are, on average, delocalized over three or four sites \textit{before} optimization as this allows localized eigenstate blocks (described in the main text) to form during optimization.
    }
    \label{fig:dE-and-J}
\end{figure}

In this section, we investigate how changes in both the overall energy gradient (given by $N \cdot \delta E$) and the inter-site hopping strength $J$ affect the results presented above. Variations in $J$ will have two primary effects on our system: firstly, an increase in $J$ will cause the Hamiltonian eigenstates to become more delocalised, and secondly, it will also increase the magnitude of the separation between each of the eigenenergies. Since the brightness of each eigenstate and the rates of the vibrational transitions between eigenstate both depend on the localization properties of the states, we would expect that varying $J$ will have a significant effect on our results. Indeed, as shown in Fig.~\ref{fig:dE-and-J}a, there does exist an optimal value at which the enhancement over a simple linear gradient is maximal however, this peak is relatively broad and significant enhancements are still possible at both smaller and larger coupling strengths. The peak enhancement occurs when the eigenstates are, on average, delocalised across three or four site in the linear energy gradient configuration (i.e.~before optimization) as indicated by the eigenstate participation ratio (PR). This makes intuitive sense given the formation of blocks of eigenstates delocalised over $\sim$ 4 sites in the `spiked' energy configurations (Fig.~\ref{fig:example-opt}), since having sufficiently delocalised eigenstates before optimization will make it easier to form the appropriate eigenstate blocks by adding energy spikes at certain sites.

Similarly, varying the overall energy gradient in the chain will also affect the localization properties of the eigenstates, with a lower gradient leading to more delocalised states. As explained in Sec.~\ref{sec:model}, in order to maximise the efficiency of the benchmark linear gradient, variations in $\delta E$ are linked to variations in the phonon spectrum structure. A byproduct of this is that $\omega_{max}$ (i.e.~the frequency at which phonon transitions are maximal) is also altered with the energy gradient. Fig.~\ref{fig:dE-and-J}b demonstrates that there also exists an optimal energy gradient at which the spiked structures are most effective; however, as before, significant enhancements are still possible away from this optimal value. The reason for poorer enhancement at steeper gradients is that, despite the Ohmic scaling in the phonon spectrum (Eq.~\eqref{eq:Sw-DL}) which increases the overall rate at which excitations are funnelled towards the EC, the system eigenstates become too localized to support any bright-dark separation of eigenstates and, therefore, exciton recombination rates are too high to allow successful transport.

To conclude this section, it is worth noting that the energetic structure of the system governed by Eq.~\eqref{eq:H-chain} only depends, up to a scale factor, on the ratio $J/\delta E$. Therefore, with a suitable re-scaling of time and environmental noise rates, all our results could instead be expressed in terms of this ratio alone. This reflects the more general applicability of our findings and underlines the fact that the mechanism we describe is not limited to the specific choice of parameters given in Table~\ref{table:param-values}.  However, for easier interpretation and more direct physical relevance, we have instead focused on varying the parameters $J$ and $\delta E$ independently in Fig.~\ref{fig:dE-and-J}, without correspondingly rescaling the environment parameters (other than the aforementioned change in $\omega_{max}$). In any case, Fig.~\ref{fig:dE-and-J} shows that within our model a biased chain is indeed always preferable to a flat one ($\delta E = 0$), as is further discussed in Appendix~\ref{apdx:zero-bias}.

\section{Variations in Environments}
\label{sec:var-envs}

\subsection{Vibrational Environment}
\label{sec:var-Sw-DL}

As well as robustness to changes in the system parameters explored above, it is also informative to investigate the effects of explicitly altering the vibrational and electromagnetic environments in which our chain system resides. In the vibrational frequency spectrum of Eq.~\eqref{eq:Sw-DL}, there are four main parameters which characterize the phonon bath, namely: \textit{i)} $\gamma_{phonon}$ -- the overall system-phonon coupling, \textit{ii)} $\omega_0$ -- the value of the high frequency cutoff, \textit{iii)} $\Gamma$ -- the width of the spectrum  and \textit{iv)} $T_{ph}$ -- the temperature of the phonon bath. We will analyse each of these parameters in turn in this section.

Firstly, increasing the overall phonon coupling leads to a simple and proportionate increase in the transition rates between each eigenstate pair. Since these transitions are the primary mechanism by which excitations are funnelled towards the extraction center, this leads to an increase in the rate at which funneling occurs, with a corresponding increase in output power regardless of the on-site energy configuration (assuming the rate $\gamma_{N\alpha}$ is not a bottleneck).

\begin{figure}
    \includegraphics[scale=0.34]{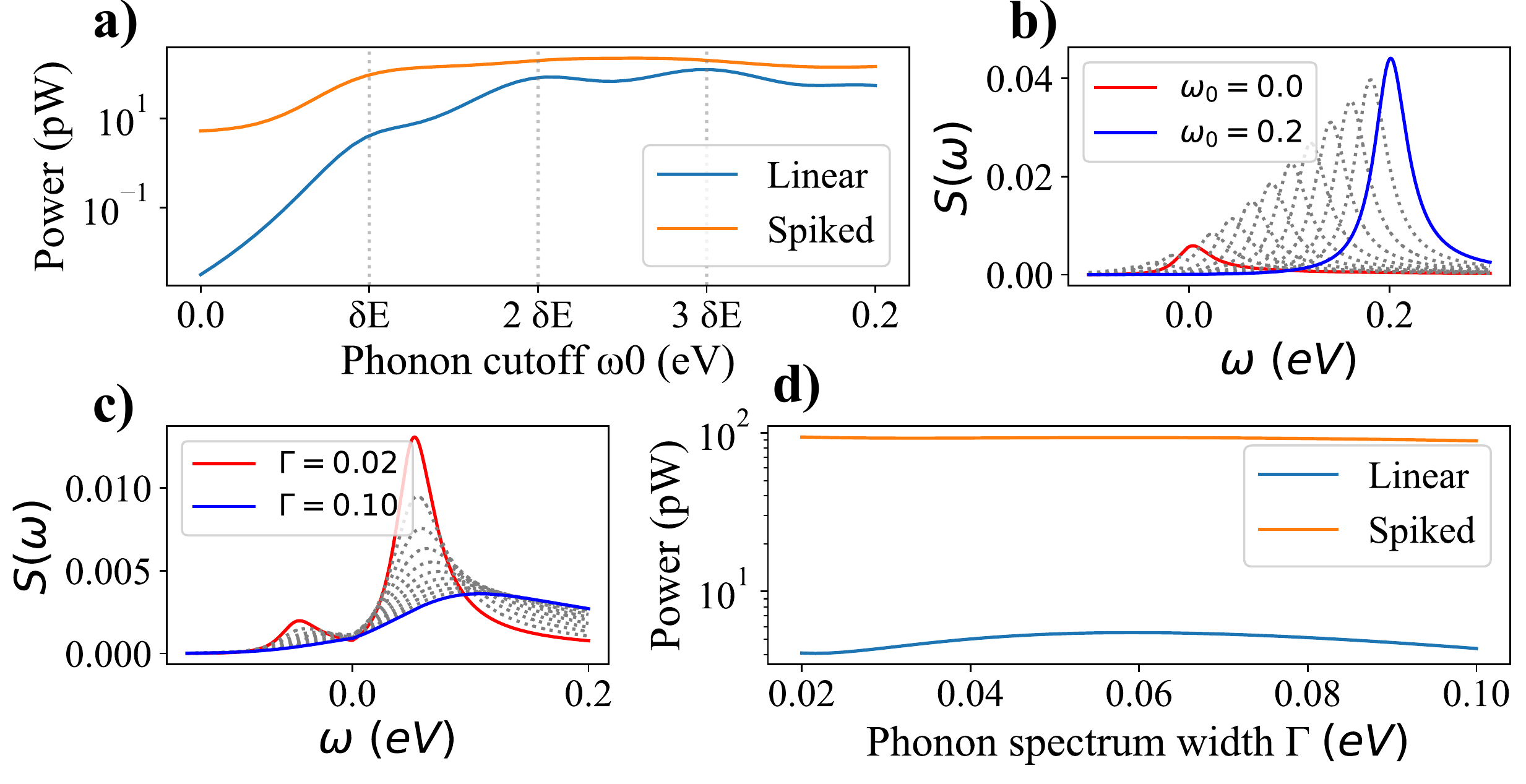}
    \caption{
    \textit{\textbf{a)}} Variation in power output vs phonon spectrum cutoff frequency for linear gradient and the optimum energy configuration shown in Fig.~\ref{fig:example-opt}. Power output from linear case is increased when spectrum peak coincides with integer multiples of energy differences between system sites ($\delta E$ = 0.05\units{eV}) but `spiked' energy configuration is still preferred at all values of $\omega_0$ and \textit{\textbf{b)}} corresponding variation in phonon spectrum shape (Eq.~\eqref{eq:Sw-DL}) as $\omega_0$ changes. \textit{\textbf{c)}} Variation in phonon spectrum shape when spectral width parameter is varied and \textit{\textbf{d)}} corresponding change in power output from linear and `spiked' chains showing that spikes are beneficial regardless of $\Gamma$ value.
    }
    \label{fig:varying-Sw}
\end{figure}

Secondly, as shown in Fig.~\ref{fig:varying-Sw}b, varying the phenomenological cutoff frequency has the effect of shifting the frequency at which $S(\omega)$ is peaked, thereby altering which eigenstate transitions happen at the fastest rates. However, due to the term in Eq.~\eqref{eq:Sw-DL} which is linear in $\omega$, it also has the added effect of increasing the height of this peak (i.e.~the rates of all phonon transitions). Fig.~\ref{fig:varying-Sw}a shows that the novel on-site energy landscape, containing regularly repeating `spikes' in energy, outperforms the linear energy gradient over a broad range of cutoff frequency values. Also highlighted, in the same figure, are the points at which this frequency matches an integer multiple of the energy difference ($\delta E$) between sites. At these points, the linear gradient system can make optimal use of phonon-mediated  transitions between pairs of system eigenstates to achieve transport but, despite these advantageous conditions, the `spiked' energy landscape proves to be more efficient.

An increase in the third free phonon spectrum parameter, $\Gamma$, leads to a broader spectrum with a lower overall peak height, both of which result in a wider range of transition frequencies having comparable rates. The bottom row of Fig.~\ref{fig:varying-Sw} shows the effect of this variation on both the spectrum itself and the resulting output powers from both linear and spiked chains. Again, we clearly see that the spiked chain is significantly better than the linear gradient configuration regardless of the phonon spectrum width and, for most spectrum widths, the addition of energy spikes provides at least an order of magnitude improvement in transport.

\begin{figure}
    \includegraphics[scale=0.47]{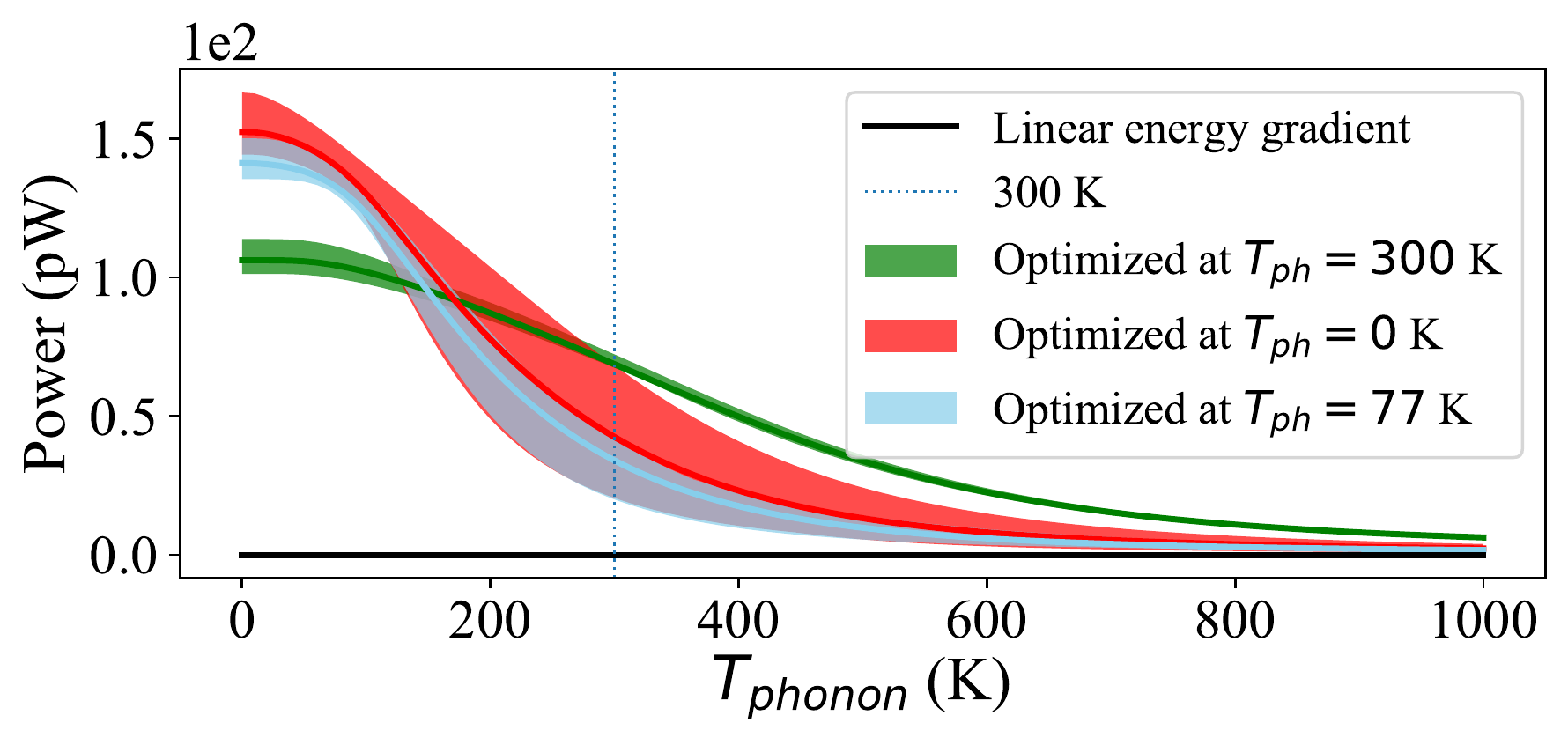}
    \caption{
    Power output vs phonon bath temperature for chains with on-site energy configurations optimized at a number of different temperatures. The power output range swept out by the best five optima found at each of the three temperatures is shown. This demonstrates that the spiked optima (Fig.~\ref{fig:example-opt}) are best at $T_{ph} \gtrsim 300\units{K}$ but that better solutions exist at low temperature. The local nature of the optimization procedure (discussed in Appendix~\ref{apdx:local-opt}) is clear here since no local optimum found at $T_{ph}=77\units{K}$ is better than the best $T_{ph}=0\units{K}$ optimum however this would not be true if the `true' global optimum was found at each $T_{ph}$.
    }
    \label{fig:varying-Tph}
\end{figure}

The final free parameter in the phonon spectrum, and indeed the most easily controllable in practice, is the temperature $T_{ph}$ of the phonon bath. A change in temperature will, in our model, primarily affect the ratio of phonon absorption vs emission events which will, in turn, dictate how quickly an excitation is funnelled down the energy gradient towards the EC. A minor secondary effect of varying $T_{ph}$ is an overall change in all phonon transition rates. Physically, a high $T_{ph}$ will lead to an increased likelihood of phonon absorption events which cause an excitation to temporarily `climb back up the chain' away from the EC. This will cause the excitons to spend more time in the chain overall, therefore increasing the chance of detrimental recombination events. Fig.~\ref{fig:varying-Tph} shows the power output vs $T_{ph}$ for chains optimized at a number of different temperatures. Here, we see that, as expected, transport is generally more efficient at lower temperatures (for the reasons just described) and that chains optimized at different temperatures are tailored to work best in those temperature regimes. In particular, and as shown in Appendix~\ref{apdx:low-Tph}, chains which are optimized at $T_{ph} = 0$\units{K} can make use of a subtly different energy configuration to the spikes in Fig.~\ref{fig:example-opt}. Specifically, since the rate of phonon absorption by the system is exactly zero, the `left-right' separation of site-basis support within each eigenstate block (described in Sec.~\ref{sec:mechanism}) is no longer necessary and it is more beneficial to create bright-dark state pairs with increased mutual site-basis support, in order to increase the magnitude of the separation in brightness between the eigenstates. This contrast in the low $T_{ph}$ limit provides further evidence that the spiked energy landscapes described in this paper are specifically optimized for realistic, finite temperature transport and are robust to moderate changes in the vibrational environment surrounding the system.

\subsection{Non-radiative loss}
\label{sec:rad-vs-nonrad}

\begin{figure}
    \includegraphics[scale=0.43]{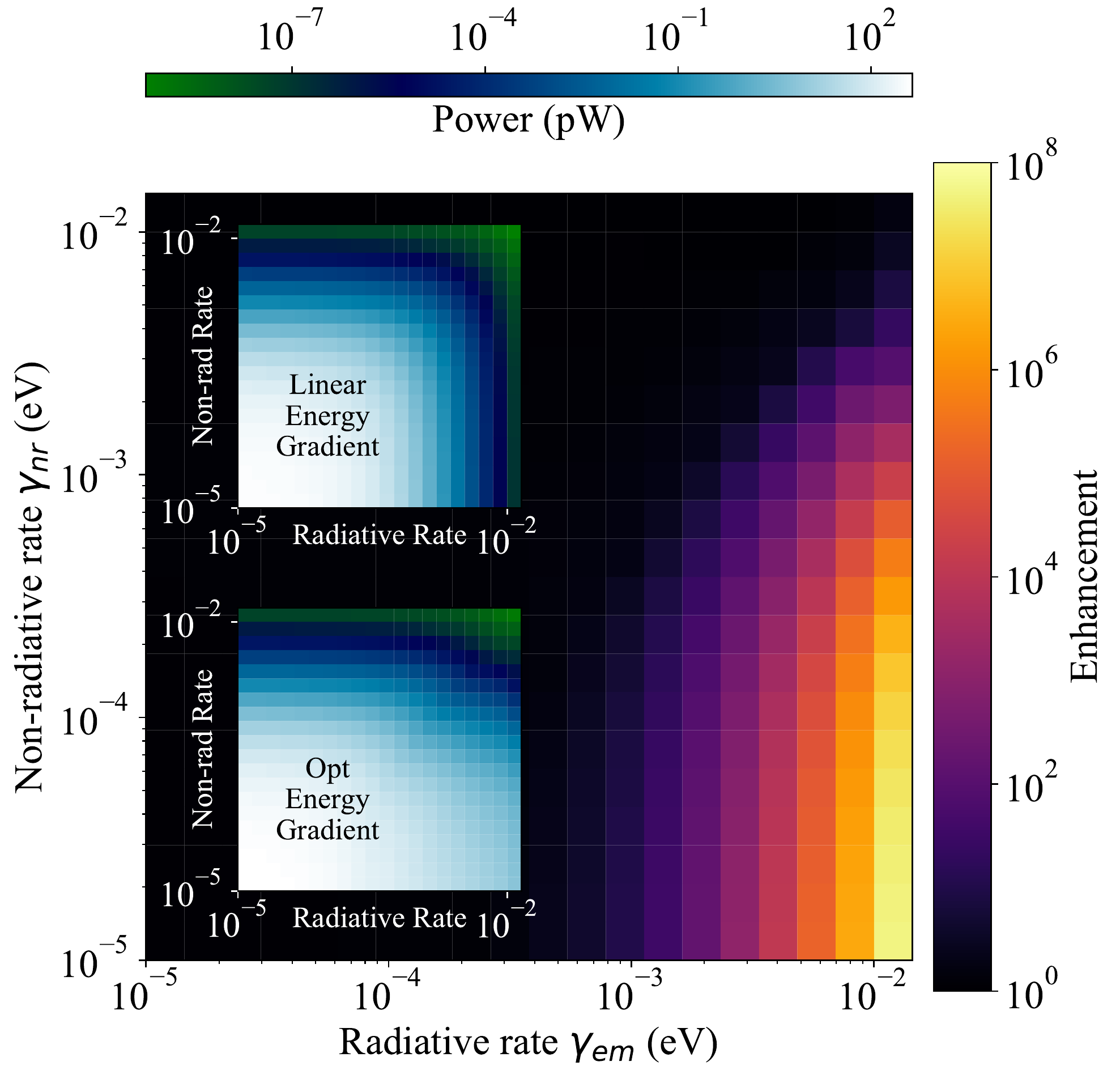}
    \caption{
    Enhancement factor vs linear gradient configuration as a function of both radiative and non-radiative decay rates demonstrating that `spiked' energy configurations are only beneficial in protecting against radiative decay processes. If non-radiative decay is dominant, radiative decay is negligible or if both decay rates are large then modifying the on-site energies fails to meaningfully improve transport. \textit{Insets} - Power outputs for both energy configurations over the same range of radiative and non-radiative decay rates.
    }
    \label{fig:nonrad-decay}
\end{figure}

In any physically realistic exciton transport system, it is highly unlikely that the collective optical emission processes used in our model thus far would be the only mechanism through which an exciton could feasibly be lost from the system. Non-radiative decay pathways involving the emission of phonons, rather than photons, will also present in practice. In order to investigate the effects of these competing decay channels in our model, rather than focusing on a specific mechanism, we simply add an additional system environment interaction which can cause transitions to the ground state, with a rate which is entirely independent of any electromagnetic environment parameters. At each individual system site $\ket{i}$, the operator for this non-radiative process takes the form
\begin{equation}
    \hat{A}_{i, nr} = \outerproduct{i}{g} + \outerproduct{g}{i}
\end{equation}
with the same flat frequency spectrum [Eq.~\eqref{eq:Sw-flat}] used previously, the same temperature $T_{cold}=300\units{K}$ and a rate denoted by $\gamma_{nr}$ which is assumed to be equal on all sites for simplicity. Due to the lack of a sum over sites (cf. Eq.~\ref{eq:optical-op}), this process involves individual decay from each system site and does not give rise to any interference effects between decays from different system sites. Therefore, the distinction between bright and dark eigenstates will have no effect on the rate at which this non-radiative process acts. Therefore, we would expect that the benefits of adding spikes in site energy, which lead to the separation of the system into dark and bright states, would be negligible. Fig.~\ref{fig:nonrad-decay} validates this conclusion by showing the power output from linear and optimized chains, as well as the relative enhancement of the optimized configurations, as a function of radiative and non-radiative decay rates ($\gamma_{em}$ and $\gamma_{nr}$ respectively). It shows that when non-radiative decay is the dominant loss channel the optimization procedure has very little effect since there is no mechanism by which altering the on-site energies can stem the flow of these losses. Conversely, when non-radiative decay is negligible and radiative (collective) decay processes are fast, the addition of energy spikes can lead to remarkable improvements (up to 7 orders of magnitude) in transport efficiency. 

As a final remark on Fig.~\ref{fig:nonrad-decay}, it also illustrates the (trivial) observation that, in our model, transport is always going to be more efficient with small loss rates of all kinds, and that  transport will be largely unsuccessful if both loss rates are high. However, it is important to note that, in a realistic system with dipole-dipole-like couplings between sites, there will exist an intrinsic link between the coherent hopping rates and the optical decay rates ($J$ and $\gamma_{em}$ in Fig.~\ref{fig:system}) which is not present in our simplified model. Therefore, a system with negligible optical decay rates would, in practice, likely also possess weaker coherent hopping interactions, with an associated detrimental effect on transport efficiency due to the resulting lack of eigenstate delocalization.

\section{`Grouped' Optimization of Longer Chains}
\label{sec:group-opt}

Having gained an understanding of why these on-site energy spikes are beneficial for transport, and demonstrated the robustness of these benefits to changes in various system and environment parameters, we now investigate the possibility of utilizing the results presented above as a building block structure to enable transport over significantly longer chains. To do this, we use a modified optimization procedure which focuses on finding the optimal energy configuration for a small subsection of the chain (of size $N_\text{group}$ sites) and then copies this sub-section structure along the length of the chain, with appropriate energy shifts between each group in order to maintain an overall energy gradient. (See Appendix~\ref{apdx:group-opt} for a detailed description of this procedure.) This method makes the optimization of much longer chains numerically feasible due to the significantly reduced number of independent parameters that need to be optimized.

\begin{figure}
    \includegraphics[scale=0.39]{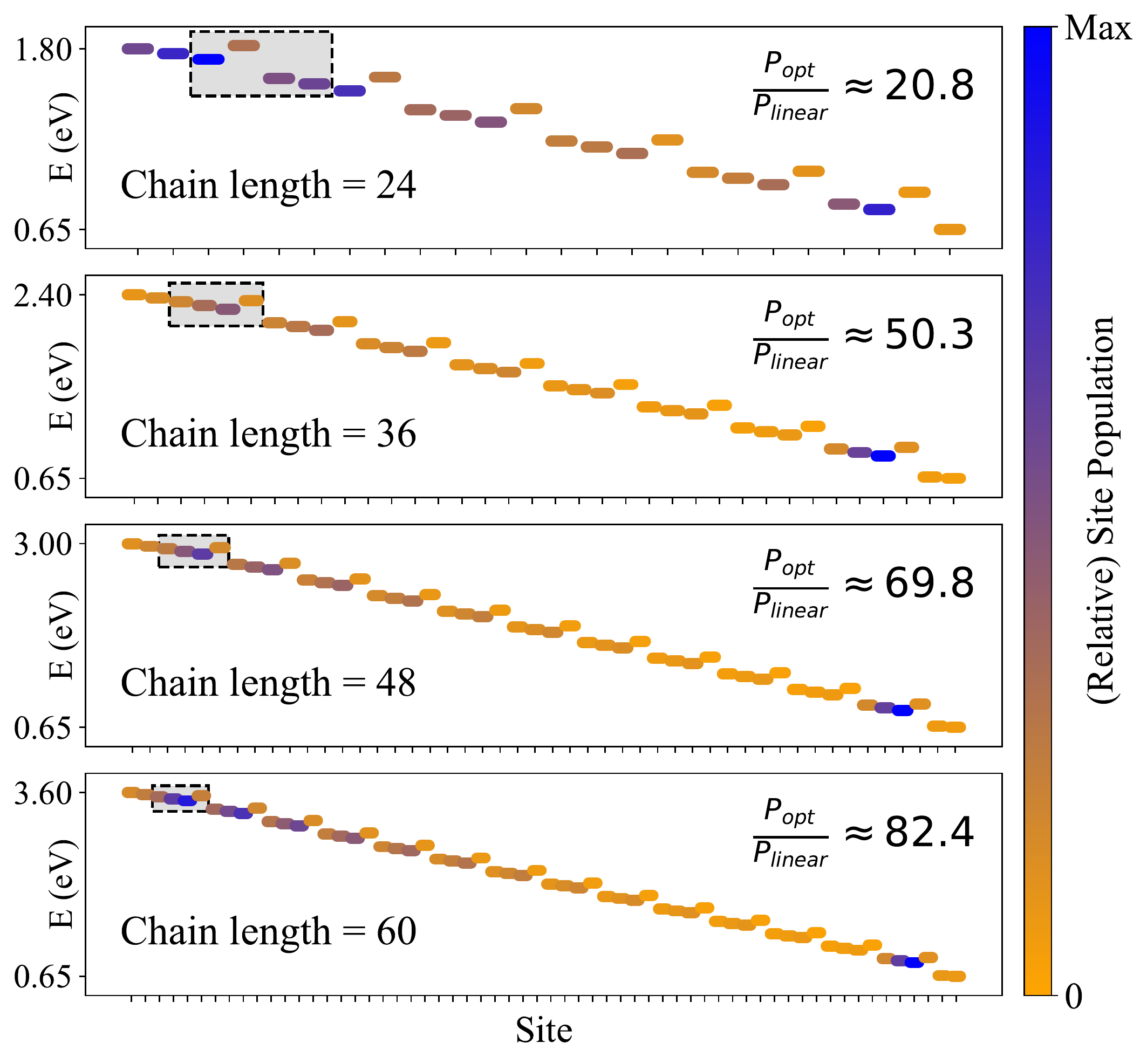}
    \caption{
    Use of regular on-site energy spikes as repeatable building block for efficient transport through long chains. In all chains the nearest neighbour coupling between sites is set at $J=0.1 \units{eV}$, the final on-site energy in the chain is $\varepsilon_N = 0.65 \units{eV}$ (therefore $\varepsilon_1 = \varepsilon_N + N \cdot \delta E$) and the energy gradient between neighbouring sites (before optimization) is fixed at $\delta E = 0.05 \units{eV}$. The group of site energies which are optimized and then repeated down the chain (with energy shift $N_\text{group} \cdot \delta E = 0.2 \units{eV}$ between groups) is highlighted in each chain.
    }
    \label{fig:grouped-optim}
\end{figure}

Using this modified approach, we find that it is indeed possible to use identically repeated groups of optimized on-site energies as a building block structure to significantly enhance transport through much longer chains. Fig.~\ref{fig:grouped-optim} demonstrates this for chains up to length $N=60$ sites, where the highlighted group of four sites is repeated (14 times for the $N=60$ case) along the energy gradient. The first and last on-site energies are fixed, and the second and penultimate energies are independently optimized to account for boundary effects. To allow for a fair comparison between chains of different length, all chains in Fig.~\ref{fig:grouped-optim} have the same total coupling between system and ambient electromagnetic field (given by $\sum_j \gamma_{em, j}$) as the 20 site chain in Fig.~\ref{fig:example-opt}. This is achieved by setting the couplings for each site to $\gamma_{em, j} = (20/N) \gamma_{0}$, where $N$ is the chain length, and $\gamma_{0} = 0.001\units{eV}$. Despite this compensation, and as shown in Fig.~\ref{fig:grouped-optim}, the enhancement ($P_{opt}/P_{linear}$) over the linear energy gradient increases with chain length. This occurs because, for transport through longer chains, the excitation necessarily has to spend more time in the chain (leading to a higher chance of radiative recombination). Therefore, adding spikes has a larger influence on radiative loss rates and is more conducive to successful transport. Even with these optimized energy configurations, the overall power output from the EC decreases monotonically with increasing chain length (again, due to increased loss via radiative recombination) which, due to numerical accuracy limitations, prevents the optimization of even longer chains. However, by repeating these optimized group structures, significant enhancements in transport efficiency relative to the linear energy gradient should, in principle, be possible for arbitrarily long chains.

In this section, we have focused on the case where $N_\text{group} = 4$, since this is the minimum group size which supports the required eigenstate block properties (i.e.~energetic and spatial separation - see Sec.~\ref{sec:mechanism}) for energy spikes to enhance transport. In Appendix~\ref{apdx:group-opt}, we show that similar results are also obtained with larger group sizes of five or six sites. However, in any practical implementation it would likely be desirable to use the smallest group size possible.

Experimentally, engineered  molecular wires exist for both charge and energy transport, for instance based on chains of porphyrins,\cite{PorphyrinDimer,LongrangePorphyrin} but also utilising more heterogeneous arrangements.\cite{Kodis, Vura-Weis1547} This raises the prospect of realising a molecular block of length four or five sites and tuning its energetic landscape and offset using gate electrodes.

\section{Conclusion}
\label{sec:conclusion}

To summarise, in this paper, we have constructed a model to describe excitonic energy transport along a chain of sites with an intrinsic energy gradient and which is coupled to ambient electromagnetic and vibrational environments. By performing a numerical optimization of the on-site excitation energies in the chain, we have shown that a linear energy gradient is far from optimal for transport, and that significant enhancements in transport efficiency can be realised by introducing regularly spaced energetic barriers (or energy `spikes'). The underlying mechanisms which facilitate this enhancement rely on a separation of the system eigenstates into optically bright and dark states and utilize phonon-mediated transitions between these eigenstates to allow transport to proceed primarily through the dark states in the system. This leads to markedly reduced energy loss from radiative exciton recombination events, resulting in successful transport occurring more often.

Despite the fundamental reliance of this novel transport mechanism on the delicate interplay between the energy landscape of the system and the spectra of its electromagnetic and vibrational environments, we have shown that the `spiked' on-site energy configurations are beneficial for transport over a range of parameter regimes. As a limitation of the results presented here, we have noted that these effects are reliant on collective radiative decay being the dominant loss channel during transport and, as a result, these novel energy landscapes are less effective if exciton decay is primarily non-radiative. 

In the various Appendices to this paper we explore a number of generalisations and modifications to our model such as: distance-dependent (rather than nearest neighbour only) coherent couplings, steady-state current as an alternative measure of efficiency, optimized energy configurations at zero temperature and zero energy bias, and limitations on the exciton extraction rate $\gamma_{N\alpha}$. None of the results presented therein differ significantly from the optimum energy configurations and transport mechanisms discussed throughout the main text.

Finally, we made use of the regular nature of the optimal energy configurations to propose their use as a building block for designing structures which improve transport efficiency through much longer chains. An interesting area for future exploration would be to investigate whether similar effects can benefit transport in higher dimensions, e.g.~along membranes. For instance, we expect similar enhancements to be available for planar arrangements of rings similar to certain photosynthetic structures, whereas applicability to arbitrary 2D or 3D systems remains an open question. Extending the concepts developed here to such structures could potentially help to mitigate the restrictive nature of exciton diffusion lengths present in many organic photo-voltaic devices and in nanoscale energy distribution more generally.

\section*{Supplementary Material}

See \href{https://aip.scitation.org/doi/suppl/10.1063/5.0023702}{supplementary material} for additional versions of Fig.~\ref{fig:param-sens},~\ref{fig:example-opt},~\ref{fig:eigenbasis},~\ref{fig:dE-and-J}~\&~\ref{fig:varying-Sw} which were produced using a Bloch-Redfield master equation instead of the PME used in the main text.

\begin{acknowledgments}
We thank Simon Benjamin for insightful discussions. SD was supported by the EPSRC Grant No.~EP/L015110/1. AF thanks the Anglo-Israeli and the Anglo-Jewish associations, and the Leverhulme Trust (RPG-080) for funding. This work was further supported by EPSRC Grants No. EP/T007214/1 and EP/T01377X/1.
Computations were performed using the open source QuantumOptics.jl~\cite{QO.jl} and Optim.jl~\cite{Optim.jl} packages.
The data that support the findings of this study are available from the corresponding author upon reasonable request.
\end{acknowledgments}

\appendix

\section{Distance-Dependent Couplings}
\label{apdx:dist-dep}

While the model in the main text uses nearest neighbour only coupling in the system Hamiltonian to simplify the analysis, in this Appendix we show that the novel energy optima describe throughout this paper are not exclusive to this form of coherent coupling. To this end, we adopt a form of distance-dependent coupling here in which inter-site coupling strengths are given by
\begin{equation}
    D(\vec{r}_i, \vec{r}_j) = \frac{J}{2|\vec{r}_i - \vec{r}_j|^3} \, ,
\end{equation}
where $\vec{r}_{i}$ is the real space position of site $\ket{i}$ and $J$ is an overall coupling constant. To allow for a fair comparison with Fig.~\ref{fig:example-opt} in the main text, we set $J=0.1\units{eV}$ and the co-ordinates of the $i$-th site in the chain as $(i, 0, 0)$ (in arbitrary units). This provides nearest neighbour hopping of the same strength as used in Fig.~\ref{fig:example-opt} so that the only difference is the additional weaker pair-wise coupling between other sites which are further apart.

\begin{figure}
    \includegraphics[scale=0.43]{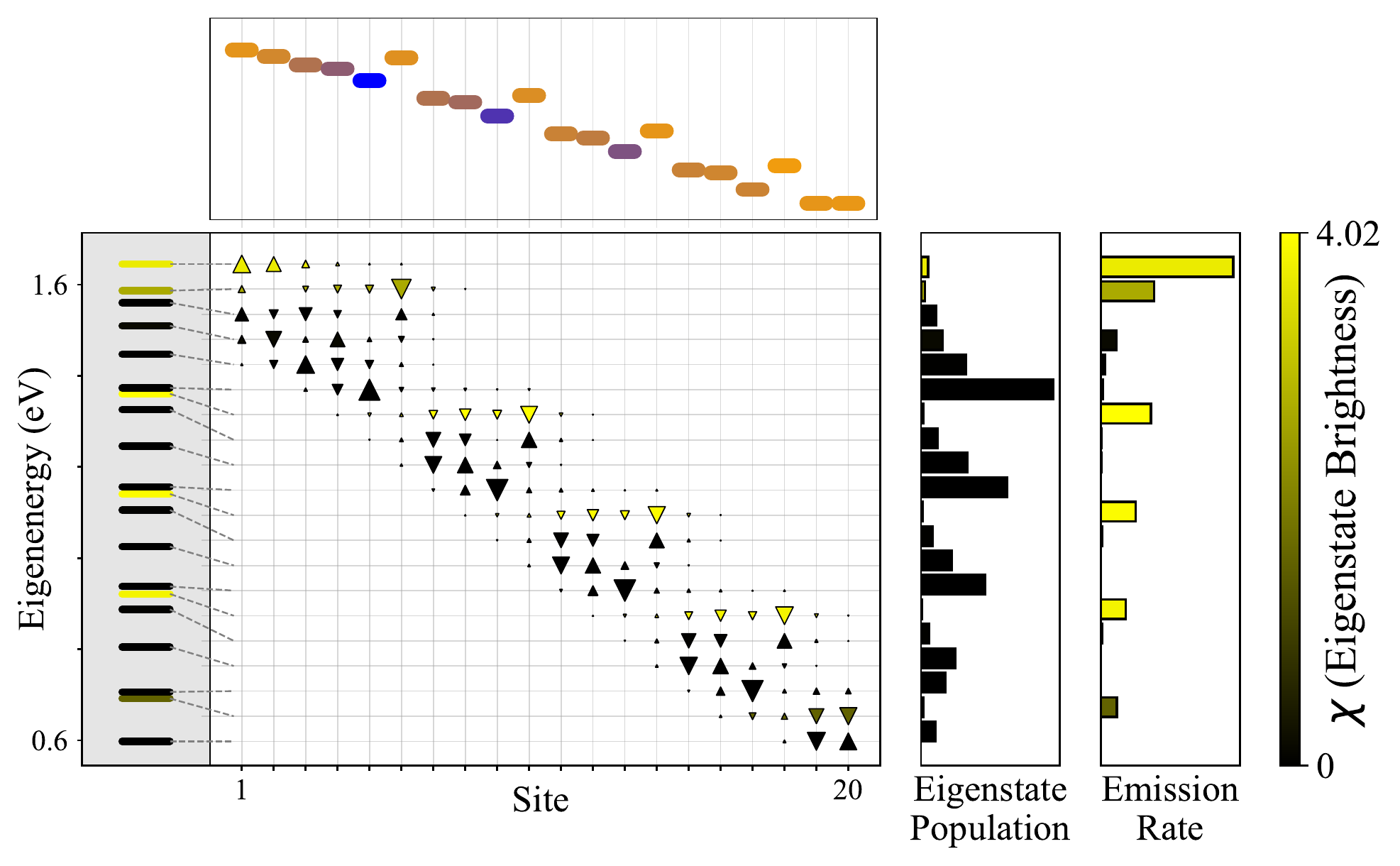}
    \caption{
    \textit{Top} A typical optimum energy landscape found for a model with fully distance-dependent inter-site coupling. \textit{Bottom} - The associated eigenbasis properties for this energy landscape with the same block structure in terms of eigenstate localization and bright/dark state separation as examined in the main text (i.e.~Fig.~\ref{fig:eigenbasis}).
    }
    \label{fig:dist-dep}
\end{figure}

Fig.~\ref{fig:dist-dep} shows a typical optimization result using this modified coherent coupling and clearly demonstrates that the benefits to transport of adding regularly spaced on-site energy spikes also hold with this more realistic coupling model.

\section{Steady State Current}
\label{apdx:ss-current}

Due to the fact that the novel energy configurations found in the main text primarily aid transport by reducing undesirable losses from the system, it would also be expected that these configurations benefit other measures of transport efficiency. In this section we explicitly verify this expectation by using the steady state \textit{current} as an alternate measure of transport efficiency. We define this exciton current as
\begin{equation}
    I_\text{out} = \gamma_{N\alpha} P_N ~,
\end{equation}
where $\gamma_{N\alpha}$ is the extraction rate from the last site $\ket{N}$ of the chain (see Fig.~\ref{fig:system}) which has population $P_N$. In the main text, the open quantum processes which allowed excitons to flow through the `extraction center' (EC) were bi-directional processes with a net energy flow out of the system due to finite temperature considerations. However, in the steady state current case, these two-way processes can artificially inflate the current output from the system by inadvertently funneling excitations from the extraction center/ground state directly to site $\ket{N}$ without those excitations having been transported through the chain. To avoid this unwanted contribution, we set the temperature of these processes to $T_\text{cold} = 0\units{K}$ so they become truly one-way processes (while leaving the local phonon bath temperature on each system site at $T_{ph} = 300\units{K}$). Additionally, in order to avoid the EC becoming a bottleneck for this model, we also increase the rates $\gamma_{\alpha\beta}$ and $\gamma_{\beta g}$ so as to maximize flow through the EC. These changes to temperatures and rates lead to the effective removal of the EC from the system and make our model equivalent to a simple $N$ site plus ground state system which is usually used for calculating steady state currents.

\begin{figure}
    \includegraphics[scale=0.43]{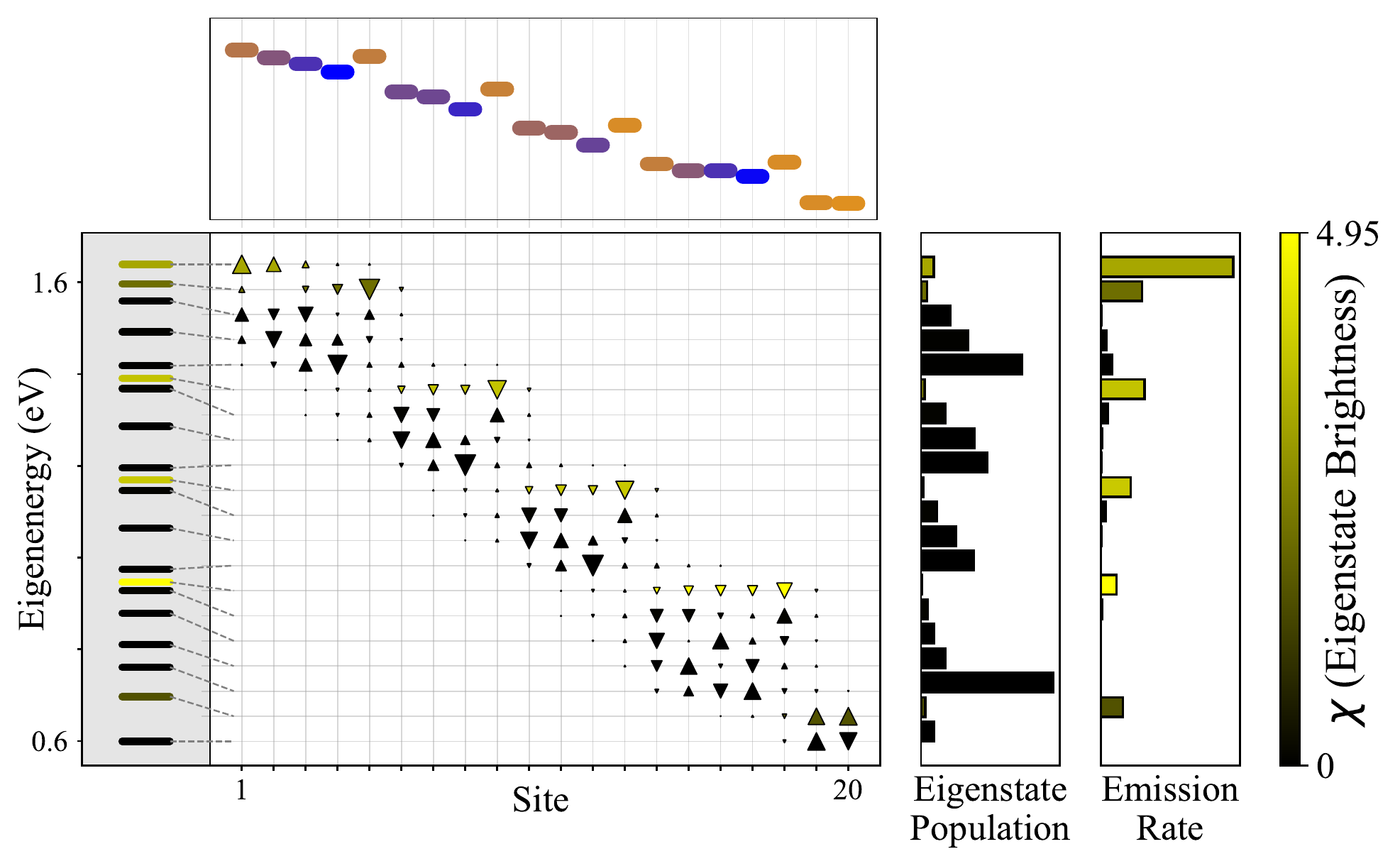}
    \caption{
    \textit{Top} - Optimal energy landscape and - \textit{bottom} - the resultant eigenbasis properties when the biased chain system is optimized for maximal steady state \textit{current} instead of power. We see close agreement between this figure and the results presented in the main text (i.e.~Fig~\ref{fig:eigenbasis}).
    }
    \label{fig:ss-current}
\end{figure}

After accounting for these modifications, we then optimize the on-site energy configuration of a 20 site chain in order to maximize this steady state current and, as shown in Fig.~\ref{fig:ss-current}, we find very similar optimal energy configurations to those in the main text. Specifically, we see the same distinctive eigenstate block structure with groups of eigenstates localized over four or five system sites, a minimally populated bright state at the top of each block and most of the population in the lower energy dark states within each block. This leads us to conclude that the benefits of adding regularly repeating on-site energy spikes to these biased chains are not exclusive to steady state power output and that these configurations are in fact generally favourable whenever detrimental radiative decay processes are present.

\section{Parameter Labels in Fig.~\ref{fig:param-sens}}
\label{apdx:param-table}

For completeness, an exhaustive list of the parameters included in the sensitivity analysis of Sec.~\ref{sec:param-sens} (and Fig.~\ref{fig:param-sens} specifically) is shown in Table~\ref{tab:param-sens-map} along with the associated index for each parameter in the x-axis of Fig.~\ref{fig:param-sens}. The description column states the points in the main text where each parameter is introduced and defined.

\begin{table}
    \centering
    \begin{tabular}{| c | c | c |}
        \hline
        \textbf{Index} & \textbf{Parameter} & \textbf{Description} \\ 
        \hline
        1 - 18 & $\varepsilon_{2-19}$ & On-site energies to optimize \\ 
        19, 20 & $\beta_{\alpha\beta}$, $\gamma_{\alpha\beta}$ & $\ket{\alpha} \leftrightarrow \ket{\beta}$ process (Eq.~\ref{eq:Sw-flat})  \\ 
        21, 22 & $\beta_{\beta g}$, $\gamma_{\beta g}$ & $\ket{\beta} \leftrightarrow \ket{g}$ process (Eq.~\ref{eq:Sw-flat},~\ref{eq:generic-env-op}) \\ 
        23-26 & $\beta_{ph, 1}$, $\Gamma_{1}$, $\omega_{0, 1}$, $\gamma_{ph, 1}$  & Site 1 phonon bath (Eq.~\ref{eq:Sw-flat},~\ref{eq:phonon-op}) \\ 
        27-30 & $\beta_{ph, 2}$, $\Gamma_{2}$, $\omega_{0, 2}$, $\gamma_{ph, 2}$  & Site 2 phonon bath (Eq.~\ref{eq:Sw-flat},~\ref{eq:phonon-op}) \\ 
        31-34 & $\beta_{ph, 3}$, $\Gamma_{3}$, $\omega_{0, 3}$, $\gamma_{ph, 3}$  & Site 3 phonon bath (Eq.~\ref{eq:Sw-flat},~\ref{eq:phonon-op}) \\ 
        
        \vdots & \vdots & \vdots \\ 
        
        95-98 & $\beta_{ph, 19}$, $\Gamma_{19}$, $\omega_{0, 19}$, $\gamma_{ph, 19}$  & Site 19 phonon bath (Eq.~\ref{eq:Sw-flat},~\ref{eq:phonon-op}) \\ 
        99-102 & $\beta_{ph, 20}$, $\Gamma_{20}$, $\omega_{0, 20}$, $\gamma_{ph, 20}$  & Site 20 phonon bath (Eq.~\ref{eq:Sw-flat},~\ref{eq:phonon-op}) \\ 
        103, 104 & $\beta_{inj}$, $\gamma_{inj}$ & $\ket{g} \leftrightarrow \ket{1}$ process (Eqs.~\ref{eq:Sw-flat},~\ref{eq:inj-op}) \\
        105, 106 & $\beta_{N, \alpha}$, $\gamma_{N, \alpha}$ & $\ket{N} \leftrightarrow \ket{\alpha}$ process (Eq.~\ref{eq:Sw-flat},~\ref{eq:generic-env-op}) \\
        107, 108 & $\beta_{em}$, $\gamma_{em}$ & $\ket{N} \leftrightarrow \ket{\alpha}$ process (Eq.~\ref{eq:Sw-flat},~\ref{eq:optical-op}) \\
        109 & $J$ & Inter-site coupling (Eq.~\ref{eq:H-chain}) \\
        \hline
    \end{tabular}
    \caption{List of all model parameters included in the parameter sensitivity analysis in the main text. In all cases in the table, parameters $\beta_x$ are given by $\beta_x = 1 / k_b T_y$ where $y$ is the relevant temperature for process $x$ (either $T_\text{cold}$ or $T_\text{hot}$) as specified in Sec.~\ref{sec:model} of the main text.}
    \label{tab:param-sens-map}
\end{table}

\section{Prevalence of local optima}
\label{apdx:local-opt}

As mentioned in the main text, the parameter space landscape for our energy transport model is extremely complex and contains a plethora of local optima which all have similar transport performance but qualitatively different energy configurations. This problem becomes more pronounced as the number of sites in the chain, and therefore the dimension of the parameter space, is increased.

In order to deal with this abundance of local optima, we start our optimization algorithm from 100 slightly perturbed energy configurations in addition to the perfect linear energy gradient. For each of these modified starting points we add a small amount of independent random disorder (sampled from a uniform distribution on [-0.01, 0.01]) to each of the on-site energies in the chain (excluding $\varepsilon_{1}$ \& $\varepsilon_N$) before commencing the numerical optimization. When carrying out the optimization itself, three separate optimization approaches were employed, namely; a simple Nelder-Mead algorithm, a simple BFGS algorithm (see [\onlinecite{Optim.jl}] for implementation details of both algorithms) and a combined, sequential method in which the BFGS method was employed \textit{after} the Nelder-Mead method, with the starting point for the BFGS method taken to be the optimum energy configuration calculated via Nelder-Mead. Having performed the optimization from each of these independent starting energy configurations, we then rank the local optima in terms of their relative improvement over the perfect linear gradient (i.e.~without disorder) and look for common features in the best performing local optima.

\begin{figure}
    \includegraphics[scale=0.45]{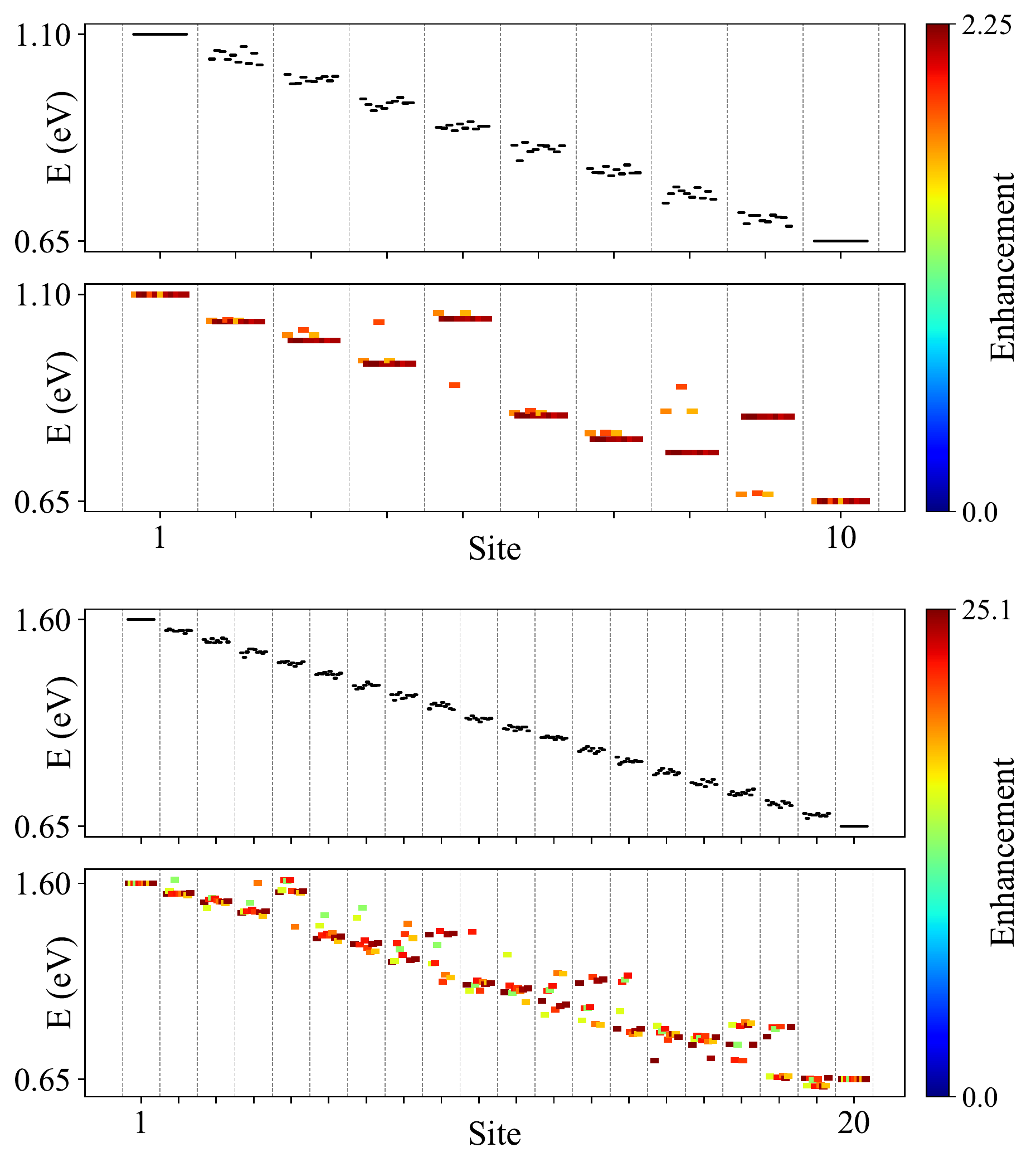}
    \caption{
    Illustration of the prevalence of local optima with increasing chain length. For each chain length, the bottom panel shows the 10 best performing configurations found from 100 starting points and the top panel shows the randomly perturbed starting configurations for each of these optima. Within each site boundary, each point (or short horizontal line) corresponds to the on-site energy of that site for a particular initial/optimal configuration. The apparent preference for energy spikes on sites 5 and 9 stems from the benefits of forming complete eigenstate block structures as close to the top of the chain as possible (since this is where the majority of the population is lost pre-optimization).
    }
    \label{fig:lots-of-optima}
\end{figure}

Fig.~\ref{fig:lots-of-optima} demonstrates the parameter space complexity scaling with chain length by comparing the best 10 optima found for both 10 and 20 site chains. In the 10 site case, we find that most of starting points have converged to the same the local optimum which has energy `spikes' on the sites 5 and 9. In contrast, for the 20 site chain we find a broader range of local optima with no obvious pattern in precisely which sites should have raised energies. Despite this, many of these different configurations exhibit at least a factor of 10 enhancement over the 20 site chain perfect linear gradient. Furthermore, although it is difficult to discern from Fig.~\ref{fig:lots-of-optima}, as explained in the main text, many of these different local optima have a similar eigenbasis structure and mechanism by which they improve transport performance.

\begin{figure}
    \includegraphics[scale=0.42]{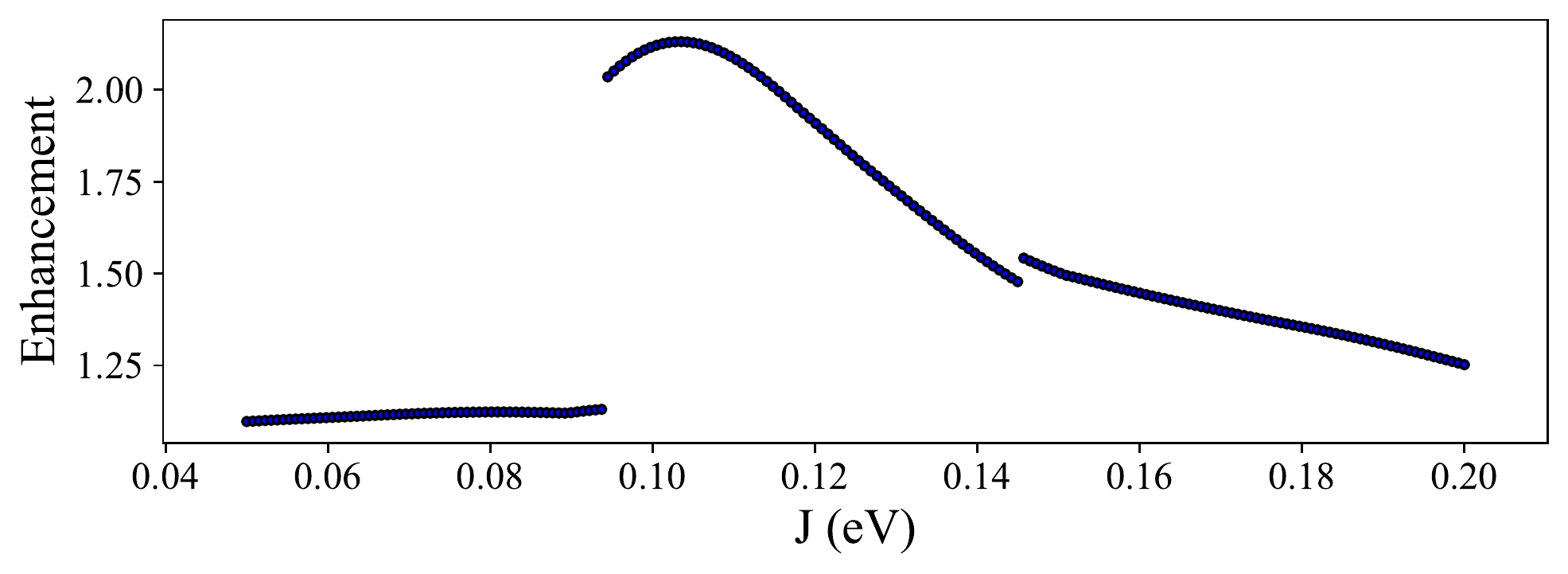}
    \caption{
    Illustration of the spurious `phase-transition' type features that can appear when using simpler numerical optimization approaches but which do not reflect any physical features within the model system. (Chain of length $N=10$ sites used here for computational convenience.)
    }
    \label{fig:fake-phase-transition}
\end{figure}

As a further illustrative example of the care required when analysing the on-site energy optimization results it is worth noting that, as originally observed in Ref.~[\onlinecite{OQS:Amir-thesis}], surprising results are obtained when using certain numerical optimization approaches. Fig.~\ref{fig:fake-phase-transition} is a replica of Fig.~\ref{fig:dE-and-J} from the main text except that, instead of the ensemble optimization method described above which uses a combination of Nelder-Mead and BFGS algorithms, a \textit{single} optimization run using the L-BFGS algorithm is performed at each $J$ value (starting from a perfect linear energy gradient). Since this (gradient-based) optimization method simply converges to the nearest local maximum in parameter space, the resulting curve looks strikingly different to the main text version and there appears to be some kind of phase-transition-like occurrence just below $J=0.1\units{eV}$ (and another smaller feature above $J=0.14\units{eV}$). However, this sharp discontinuity is an artifact of the optimization procedure and vanishes when the more robust ensemble optimization approach, used throughout the rest of this paper, is employed.

\section{Phonon rates and site basis overlap}
\label{apdx:phonon-rates}

As mentioned in Sec~\ref{sec:mechanism} of the main text, the `left-right' separation of bright/dark eigenstates in terms of site-basis localization plays a crucial role in minimizing population of the bright eigenstates and therefore reduces the rate of detrimental radiative recombination. It makes intuitive sense that interactions with a local (in the site basis) phonon bath should depend on the support of the various eigenstates on the corresponding system site. In this Appendix, we explicitly derive this dependence for a PME of the form given in Eq.~\eqref{eq:pme}.

The terms in the PME which pertain to a single system-environment interaction are determined by isolating a single term in the sum over $\mu$ in Eq.~\eqref{eq:W-mat}. Explicitly, they are given by
\begin{equation}
    \label{eq:W-term}
    W_{nm} = S_\mu (\omega_{mn}) \bra{\phi_m} A_\mu \ket{\phi_n} \bra{\phi_n} A_\mu \ket{\phi_m} ~,
\end{equation}
where $A_\mu$ is the system part of the relevant system-environment interaction and where $\phi_n$ and $\phi_m$ are the system eigenstates between which the environment is mediating interactions. In our model, for a local phonon bath coupled to the $i$-th site in the chain, we use $A_\mu = \outerproduct{i}{i}$ to describe phonon-mediated exchange of population between eigenstates. Therefore, by expanding both eigenstates in the site basis, we find
\begin{align}
   \bra{\phi_m} A_\mu \ket{\phi_n} &= \left( \sum_k c_{mk}^* \bra{k} \right) \outerproduct{i}{i} \left( \sum_l c_{nl} \ket{l} \right) \notag \\
   &= \sum_{k, l} c_{mk}^* \delta_{ki} \delta_{il} c_{nl} \notag \\
   &= c_{mi}^* c_{ni} 
\end{align}
and, substituting this final expression into Eq.~\eqref{eq:W-term}, we get
\begin{equation}
    W_{nm} = S_\mu(\omega_{mn}) |c_{mi}|^2 |c_{ni}|^2 ~,
\end{equation}
demonstrating that the phonon-mediated transition rate between eigenstates $\phi_n$ and $\phi_m$ is dependent on the mutual site-basis support of both eigenstates on system site $\ket{i}$ (and also on the eigenenergy difference through the environment spectral density $S_\mu$).

\section{Zero-Bias Limit}
\label{apdx:zero-bias}

\begin{figure}
    \includegraphics[scale=0.55]{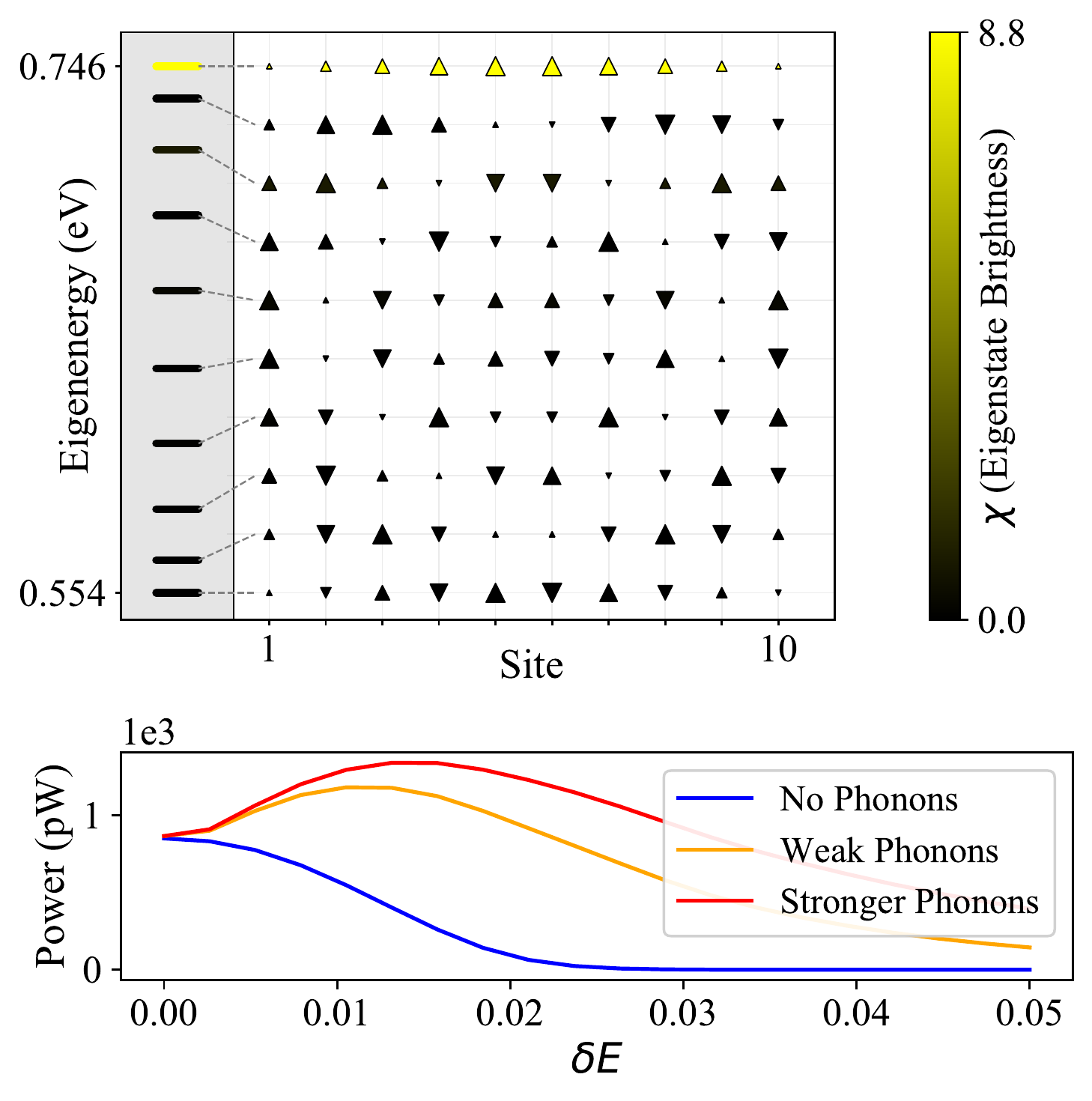}
    \caption{
    \textit{Top} - Illustration of the eigenbasis structure for the $\delta E = 0$ limit of the transport model described in the main text. \textit{Bottom} - Power vs energy gradient curves for three different system-phonon coupling strengths showing that the flat chain is only preferable when there is no system-phonon coupling.
    }
    \label{fig:zero-bias}
\end{figure}

Having focused exclusively on chains with an intrinsic energy gradient in the main text, in this Appendix we briefly analyse the $\delta E$ = 0 limit. It is well documented that a flat chain with non-zero nearest neighbour hopping ($J$) and periodic boundary conditions (i.e.~the tight binding model) has Hamiltonian eigenstates which are completely delocalised over the whole chain (Bloch-wave states). Assuming $J$ is positive, the maximally symmetric state will be highest in energy and, in the presence of an ambient electromagnetic environment, this state will also have the strongest coupling to the field (i.e it will be brightest). The eigenstate structure for a flat, finite length chain (i.e.~without periodic boundary conditions) is qualitatively similar, as shown in the top panel of Fig.~\ref{fig:zero-bias}, with some minor differences at the chain edges.

When thinking about transporting excitons in such a system, it is preferable to have as much population as possible in the eigenstates in the middle of the energy spectrum, since these eigenstates have the largest overlaps with the extraction site $\ket{N}$. However, when such a system is coupled to a finite temperature vibrational environment, phonon interactions will tend to funnel population into the lower energy eigenstates which, in turn, leads to population becoming concentrated in the middle of the chain and therefore poorer transport efficiency. This is demonstrated in the bottom panel of Fig.~\ref{fig:zero-bias} which shows that, in the absence of phonons (which is unrealistic in practice), a flat chain provides the highest possible power output whereas for finite system-phonon coupling, an energy gradient can provide much higher power output.

\section{Low $T_{ph}$ limit}
\label{apdx:low-Tph}

Section~\ref{sec:var-Sw-DL} in the main text mentions that the optimum on-site energy configurations found at low phonon bath temperatures contain slightly different features from the `spiked' energy landscapes found at $T_{ph} \gtrsim 300 \units{K}$. There exist practical systems, such as coupled quantum dots~\cite{QT:low-Temp-exciton-drift-in-QD, QT:energy-grad-indirect-QD-excitons}, to which our model may be applicable, and which operate at these lower temperatures. Therefore, in this section we briefly explain how these low temperature optimization results differ and the physical considerations which give rise to this difference.

\begin{figure}
    \includegraphics[scale=0.43]{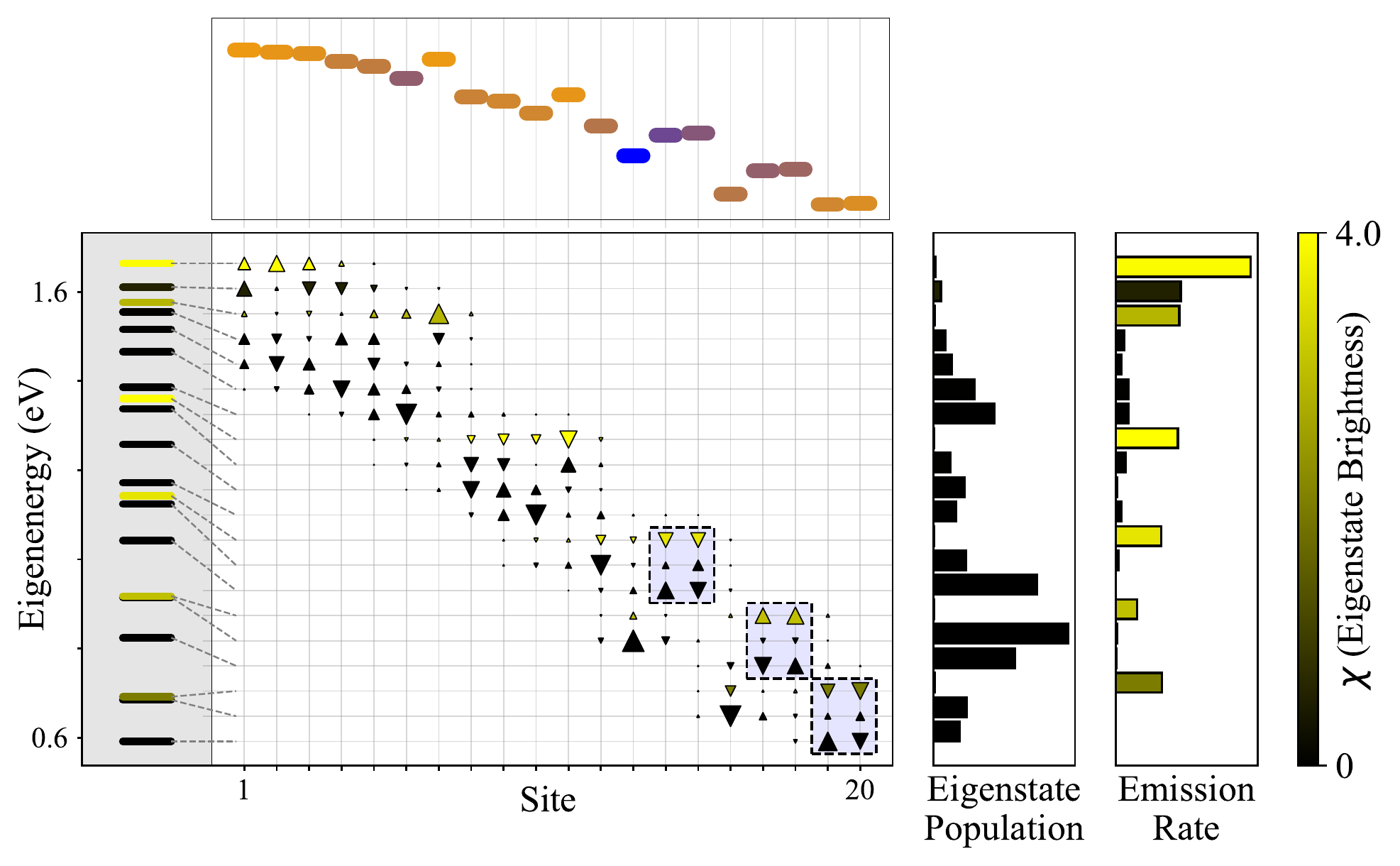}
    \caption{
    Illustration of a typical on-site energy optimization result with $T_{ph} = 0 \units{K}$. The highlighted boxes show the `dimer-like' eigenstate pairs discussed in Appendix.~\ref{apdx:low-Tph} which are uniquely beneficial to systems coupled to low temperature phonon baths due to the enhanced asymmetry in phonon absorption vs emission events at these temperatures.
    }
    \label{fig:fig:zero-Tph}
\end{figure}

Figure~\ref{fig:fig:zero-Tph} shows the optimized on-energy configuration found at $T_{ph} = 0 \units{K}$ as well as the associated eigenbasis properties. Towards the bottom of the chain, rather than the discrete four-site blocks discussed in the main text, there exists a preference for `dimer-like' eigenstate pairs which are equally delocalized over just two sites and, as a result, form bright/dark state pairs. At low temperatures, the rate of phonon absorption events is significantly reduced (and is exactly zero at $T_{ph} = 0 \units{K}$), which means that the chance of exciton loss via the specific mechanism of phonon absorption to populate a bright state followed by radiative decay is much lower. Therefore, it becomes advantageous for bright and dark states to increase shared site-basis support, since this will increase the bright-dark separation (and therefore reduce the loss rate from dark states) without leading to more phonon-absorption-related losses. These dimer-like pairs become detrimental at higher phonon bath temperatures where the left-right separation of bright and dark eigenstates is more important (as discussed in Sec.~\ref{sec:mechanism} and Appendix~\ref{apdx:phonon-rates}).

\section{More on `grouped' optimization}
\label{apdx:group-opt}
In this Appendix, we elaborate on the `grouped' optimization procedure used in Sec.~\ref{sec:group-opt} and verify that the key results are not limited to groups of four sites (i.e.~energy `spikes' at every fourth site). 

\begin{figure*}
    \centering
    \includegraphics[scale=0.4]{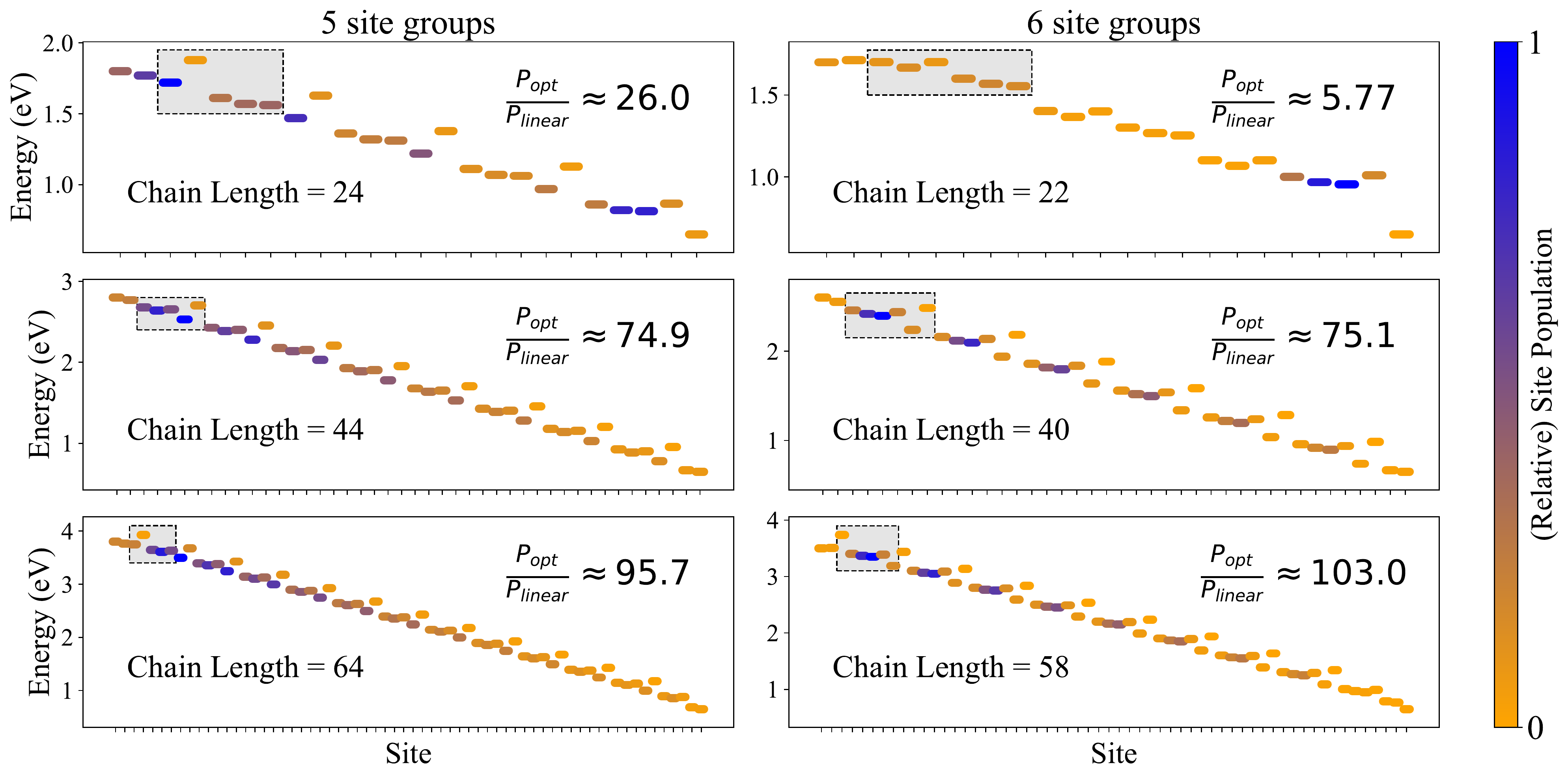}
    \caption{
    Group-based optimization of increasing chain lengths with groups of size five site (\textit{left}) and six sites (\textit{right}) respectively. The repeated group in each case is highlighted by a grey box around the relevant on-site energies.
    }
    \label{fig:groups-of-5-and-6}
\end{figure*}

As explained in Appendix~\ref{apdx:local-opt}, the optimization procedure used throughout this paper involves randomly perturbing the on-site energies in the chain before optimization. Here, we maintain this procedure. However, instead of then proceeding to optimize each of on-site energies independently, we only optimize the energies of a small subset of sites (of size $N_\text{group}$) plus a number of `edge' sites ($N_\text{edge}$) at either end of the chain (to account for boundary effects). Explicitly, with  $N_\text{edge} = 1$ and a chain of length $N$, we leave sites $1,~ N$ fixed and then optimize sites $2$ \& $N-1$ and sites $3 \rightarrow 3+N_\text{group}$ where the energies of the repeated group (sites $3$ to $3+N_\text{group}$) are then copied to sites $3+N_\text{group}+1$ to $3+2N_\text{group}$ etc. down the remainder of the chain with an energy offset of $-N_\text{group} \cdot \delta E$ between repeated groups.

For comparison with Fig.~\ref{fig:grouped-optim} in the main text, Fig.~\ref{fig:groups-of-5-and-6} shows that transport efficiency improvements are not exclusive to groups of size $N_\text{group} = 4$. For groups of both five and six sites, we find similar enhancements for a number of different chain lengths, although the `single-spike' nature of the repeated group becomes less obvious as we force the group size to become larger. This is likely due to the specific $J$ and $\delta E$ values in the chain and the resulting localization properties of the eigenstates. Regardless of these details, the significant benefits to transport efficiency are still clearly present and become more pronounced for transport through longer chains.

\section{Extraction Rate Parameter}
\label{apdx:fast-ext}

\begin{figure}
    \includegraphics[scale=0.43]{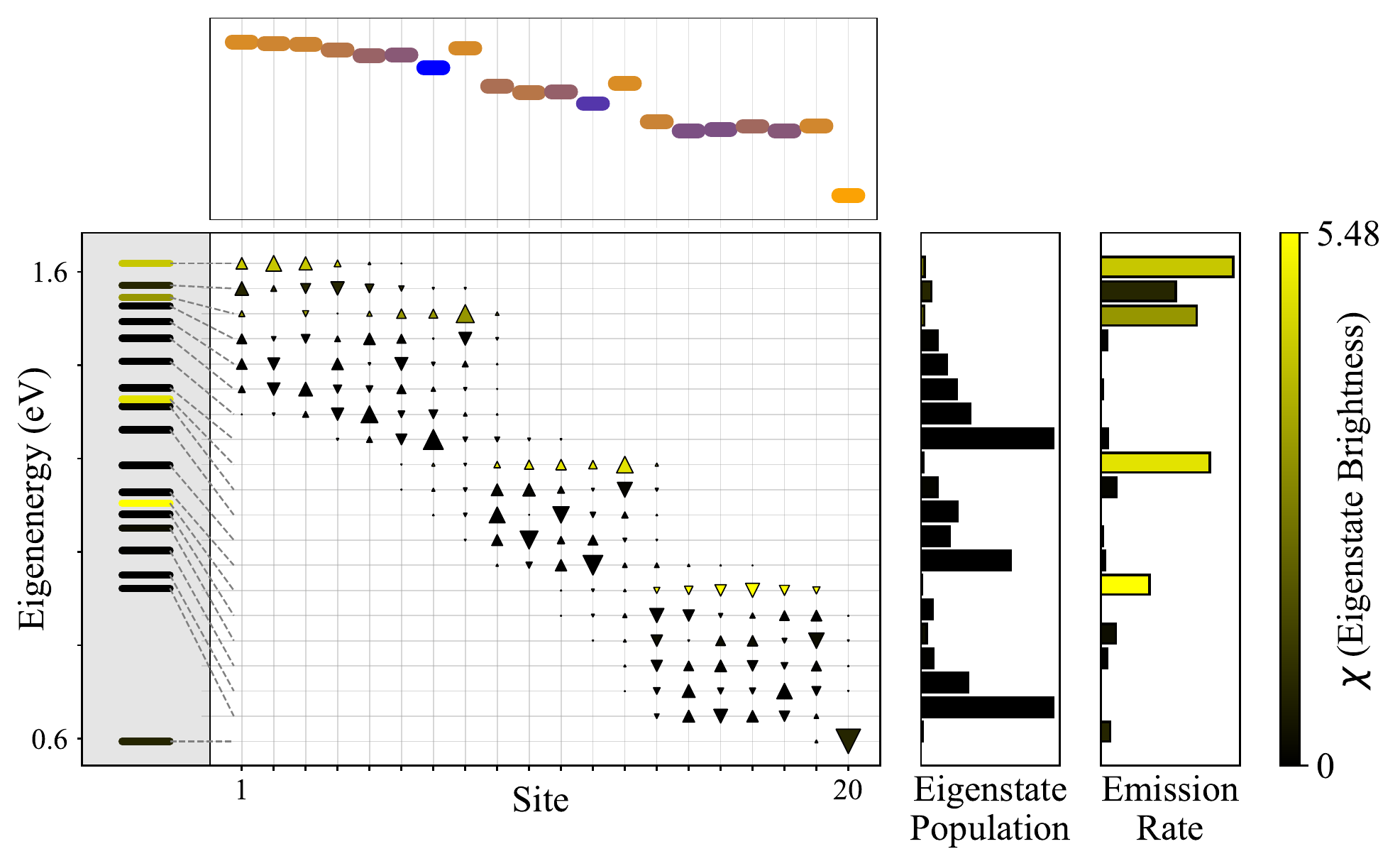}
    \caption{
    Eigenbasis properties for a typical optimization result in a system with fast extraction from the bottom of the chain. Localized eigenstate blocks are still present but there is a large gap in the eigenenergy spectrum (left-hand panel) which is unique to fast extraction rate scenarios.
    }
    \label{fig:fast-ext}
\end{figure}

Previous dark-state protection transport schemes have been found to be limited to regimes in which the extraction rate from the system ($\gamma_{N\alpha}$ in Fig.~\ref{fig:system}) is the bottleneck\cite{QT:bio-inspired-dark-state-Creatore, QT:Amir-photocell-prl, QT:Rouse-strong-dark-state-dimers, OQS:Ratchets}. In this Appendix, we show that when the extraction rate is larger than the coherent hopping rate $J$ (which is unrealistic in practice) the novel energy landscape and resulting transport through dark states remains beneficial to some degree. Figure~\ref{fig:fast-ext} shows a typical optimization result when the extraction rate $\gamma_{N\alpha}$ has been increased to $0.1 \units{eV}$ where power output at optimum is ~42 times higher than the benchmark linear energy gradient (also with fast extraction). In this case, the optimization still forms similar localized eigenstate blocks which protect again radiative recombination but we also see that a large energy gap is opened between the extraction site and the rest of the chain (by reducing the overall energy gradient in sites 1 to 19). This is beneficial when extraction is fast since it reduces the rate of phonon absorption events which move population back `up' the chain (at the cost of reduced dark-state protection in the lowest energy eigenstate) and therefore allows the unreasonably large extraction rate to be utilized most effectively. It is important to note that this is an `edge effect'.


%

\end{document}